\pdfoutput=1
\documentclass[aps,prd,twocolumn,superscriptaddress,preprintnumbers,floatfix,nofootinbib]{revtex4-2}

\usepackage{graphicx}
\usepackage{amsmath}
\usepackage[caption=false]{subfig}
\usepackage{siunitx}
\usepackage{placeins}
\usepackage{color}
\usepackage{standalone}
\usepackage{dcolumn}
\usepackage{tensor}
\usepackage{bm}
\usepackage{microtype}
\usepackage{etoolbox}
\usepackage{amssymb}
\usepackage{mathrsfs}
\usepackage{accents}
\usepackage[normalem]{ulem}
\usepackage[dvipsnames]{xcolor}
\usepackage[colorlinks,urlcolor=NavyBlue,citecolor=NavyBlue,linkcolor=NavyBlue,pdfusetitle]{hyperref}
\usepackage[all]{hypcap}
\usepackage[inline]{enumitem}
\usepackage[utf8]{inputenc}
\usepackage{xspace}
\usepackage[printonlyused, nolist]{acronym}

\AtBeginDocument{\usepackage{booktabs}}

\newcommand{\ts}{\textsuperscript}

\newcommand{\beq}{\begin{equation}}
\newcommand{\eeq}{\end{equation}}

\newcommand{\Msun}{\ensuremath{M_{\odot}}\xspace}
\newcommand{\Mtot}{\ensuremath{M_\mathrm{tot}}\xspace}

\newcommand{\chiIni}{\ensuremath{\chi_{i}}\xspace}

\newcommand{\mIni}{\ensuremath{M_{i}}\xspace}
\newcommand{\mObs}{\ensuremath{M_{\rm obs}}\xspace}
\newcommand{\tauGW}{\ensuremath{\tau_{\rm GW}}\xspace}
\newcommand{\Tage}{\ensuremath{T_{\rm age}}\xspace}
\newcommand{\fres}{\ensuremath{f_{\mathrm{res}}\xspace}}
\newcommand{\DL}{\ensuremath{D_L}\xspace}
\newcommand{\fIniLISA}{\ensuremath{f^{\rm ini}_{\rm LISA}}\xspace}
\newcommand{\fLISA}{\ensuremath{f_{\rm LISA}}\xspace}
\newcommand{\TLISA}{\ensuremath{T_{\rm LISA}}\xspace}
\newcommand{\rhoLISA}{\ensuremath{\rho_{\rm LISA}}\xspace}
\newcommand{\SLISA}{\ensuremath{S_{\rm LISA}}\xspace}

\newcommand{\fCW}{\ensuremath{f_{\rm CW}}\xspace}
\newcommand{\hthr}{\ensuremath{h_\mathrm{thr}}\xspace}

\newcommand{\mbh}{M}
\newcommand{\obh}{\Omega_{\rm BH}}
\newcommand{\obhn}{\overline{\Omega}_{\rm BH}}

\newcommand{\rg}{r_g}

\newcommand{\bosonmass}{m_b}

\newcommand{\lc}{\lambda_\mu}
\newcommand{\lbc}{\lambdabar_\mu}

\newcommand{\ob}{\omega}
\newcommand{\jb}{j}
\newcommand{\lb}{l}
\newcommand{\mb}{m}

\newcommand{\nr}{n}

\newcommand{\mc}{M_c}
\newcommand{\McompLow}{50}
\newcommand{\McompHigh}{3000}
\newcommand{\MtotLow}{100}
\newcommand{\MtotHigh}{6000}
\newcommand{\muLow}{25}
\newcommand{\muHigh}{500}
\newcommand{\fCWLow}{12}
\newcommand{\fCWHigh}{240}

\newcommand{\fgw}{f}

\newcommand{\fs}{\alpha}

\newcommand{\msun}{M_{\odot}}

\newcommand{\tinst}{\tau_{\rm inst}}

\newcommand{\nn}{\nonumber}

\newcommand*{\eq}[1]{Eq.~\eqref{eq:#1}}

\newtoggle{acrotweak}
\togglefalse{acrotweak}

\ifdefined\acrotweak
    \toggletrue{acrotweak}
\fi

\iftoggle{acrotweak}{
  
}{
}

\newtoggle{commentsoff}
\togglefalse{commentsoff}

\ifdefined\nocomments
    \toggletrue{commentsoff}
\fi

\iftoggle{commentsoff}{
  \newcommand*{\kn}[1]{}
  \newcommand*{\mi}[1]{}
  \newcommand*{\ch}[1]{}
  \newcommand*{\sv}[1]{}
  \newcommand*{\comment}[1]{}
  
  \newcommand*{\todo}[1]{}
  \newcommand*{\warn}[1]{}

}{
  \newcommand*{\kn}[1]{\textsf{\color{blue} [\textbf{KEN}: #1]}}
  \newcommand*{\mi}[1]{\textsf{\color{magenta} [\textbf{MAX}: #1]}}
  \newcommand*{\ch}[1]{\textsf{\color{RedOrange} [\textbf{CARL}: #1]}}
  \newcommand*{\sv}[1]{\textsf{\textcolor{green}{\textbf{SALVO}: #1}}}
  
  \newcommand*{\comment}[1]{\textsf{\color{blue} [\textbf{NOTE}: #1]}}
  \newcommand*{\warn}[1]{\textsf{\color{red} [\textbf{WARNING}: #1]}}
  \newcommand*{\todo}[1]{\textsf{\color{red} [\textbf{TODO}: #1]}}

}

\graphicspath{{./}}

\newcommand{\dcc}{LIGO-P2000274}

\begin{document}


\title{Multiband gravitational-wave searches for ultralight bosons}


\newcommand{\LIGOlabMIT}{\affiliation{LIGO Laboratory, Massachusetts Institute of Technology, 185 Albany St, Cambridge, Massachusetts 02139, USA}}
\newcommand{\MKI}{\affiliation{Department of Physics and Kavli Institute for Astrophysics and Space Research, Massachusetts Institute of Technology, 77 Massachusetts Ave, Cambridge, Massachusetts 02139, USA}}

\author{Ken K.~Y. Ng}
\email[]{kenkyng@mit.edu}
\LIGOlabMIT
\MKI

\author{Maximiliano Isi}
\email[]{maxisi@mit.edu}
\thanks{NHFP Einstein fellow}
\LIGOlabMIT
\MKI

\author{Carl-Johan Haster}
\LIGOlabMIT
\MKI

\author{Salvatore Vitale}
\LIGOlabMIT
\MKI

\hypersetup{pdfauthor={Ng, Isi, Haster, Vitale}}

\date{\today}

\begin{abstract}
Gravitational waves may be one of the few direct observables produced by ultralight bosons, conjectured dark matter candidates that could be the key to several problems in particle theory, high-energy physics and cosmology.
These axionlike particles could spontaneously form ``clouds'' around astrophysical black holes, leading to potent emission of continuous gravitational waves that could be detected by instruments on the ground and in space.
Although this scenario has been thoroughly studied, it has not been yet appreciated that both types of detector may be used in tandem (a practice known as ``multibanding'').
In this paper, we show that future gravitational-wave detectors on the ground and in space will be able to work together to detect ultralight bosons with masses $\muLow \lesssim \mu/\left(10^{-15}\, \mathrm{eV}\right)\lesssim \muHigh$.
In detecting binary-black-hole inspirals, the LISA space mission will provide crucial information enabling future ground-based detectors, like Cosmic Explorer or Einstein Telescope, to search for signals from boson clouds around the individual black holes in the observed binaries.
We lay out the detection strategy and, focusing on scalar bosons, chart the suitable parameter space. 
We study the impact of ignorance about the system's history, including cloud age and black hole spin.
We also consider the tidal resonances that may destroy the boson cloud before its gravitational signal becomes detectable by a ground-based follow-up.
Finally, we show how to take all of these factors into account, together with uncertainties in the LISA measurement, to obtain boson mass constraints from the ground-based observation facilitated by LISA.
\end{abstract}


\maketitle



\section{Introduction}

The existence of axions, or other ultralight bosons, could potentially solve the strong charge-parity problem~\cite{Peccei:1977hh,Peccei:1977ur,Weinberg:1977ma}, serve as stabilizing moduli in extra dimensional models~\cite{Goodsell:2009xc,Arvanitaki:2009fg}, and explain the nature of dark matter~\cite{Preskill:1982cy,Abbott:1982af,Dine:1982ah,Turner:1990uz,Hertzberg:2008wr,Jaeckel:2010ni,Essig:2013lka,Hui:2016ltb,Arvanitaki:2019rax}.
Gravitational waves (GWs) could be counted among the few observables linked to these elusive particles
\cite{Arvanitaki:2010sy,Yoshino:2013ofa,Yoshino:2014wwa,Arvanitaki:2014wva,Arvanitaki:2016qwi,Brito:2017wnc,Brito:2017zvb,Baryakhtar:2017ngi}, a realization that has motivated a flurry of research on how ground- and space-based GW detectors could join in the hunt \cite{Arvanitaki:2014wva,Arvanitaki:2016qwi,Dev:2016hxv,Brito:2017wnc,Brito:2017zvb,Baryakhtar:2017ngi,Isi:2018pzk,Ghosh:2018gaw,DAntonio:2018sff,Dergachev:2019wqa,Palomba:2019vxe,Tsukada:2018mbp,Huang:2018pbu,Ng:2019jsx}.
However, no study has yet explored the gains from simultaneously leveraging both types of GW instruments (a practice commonly known as ``multibanding'').
In this paper, we show that a coordinated use of ground- and space-based detectors will increase our chances of detecting GWs from ultralight bosons: observations of a binary inspiral signal detected by LISA \cite{LISA} will provide crucial information enabling targeted searches for ultralight bosons with third-generation (3G) GW detectors such as the \ac{ET} \cite{Punturo:2010zz} or \ac{CE} \cite{Dwyer:2014fpa, Evans:2016mbw, Reitze:2019iox}.

\begin{figure}[tb]
\centering
\includegraphics[width=\columnwidth]{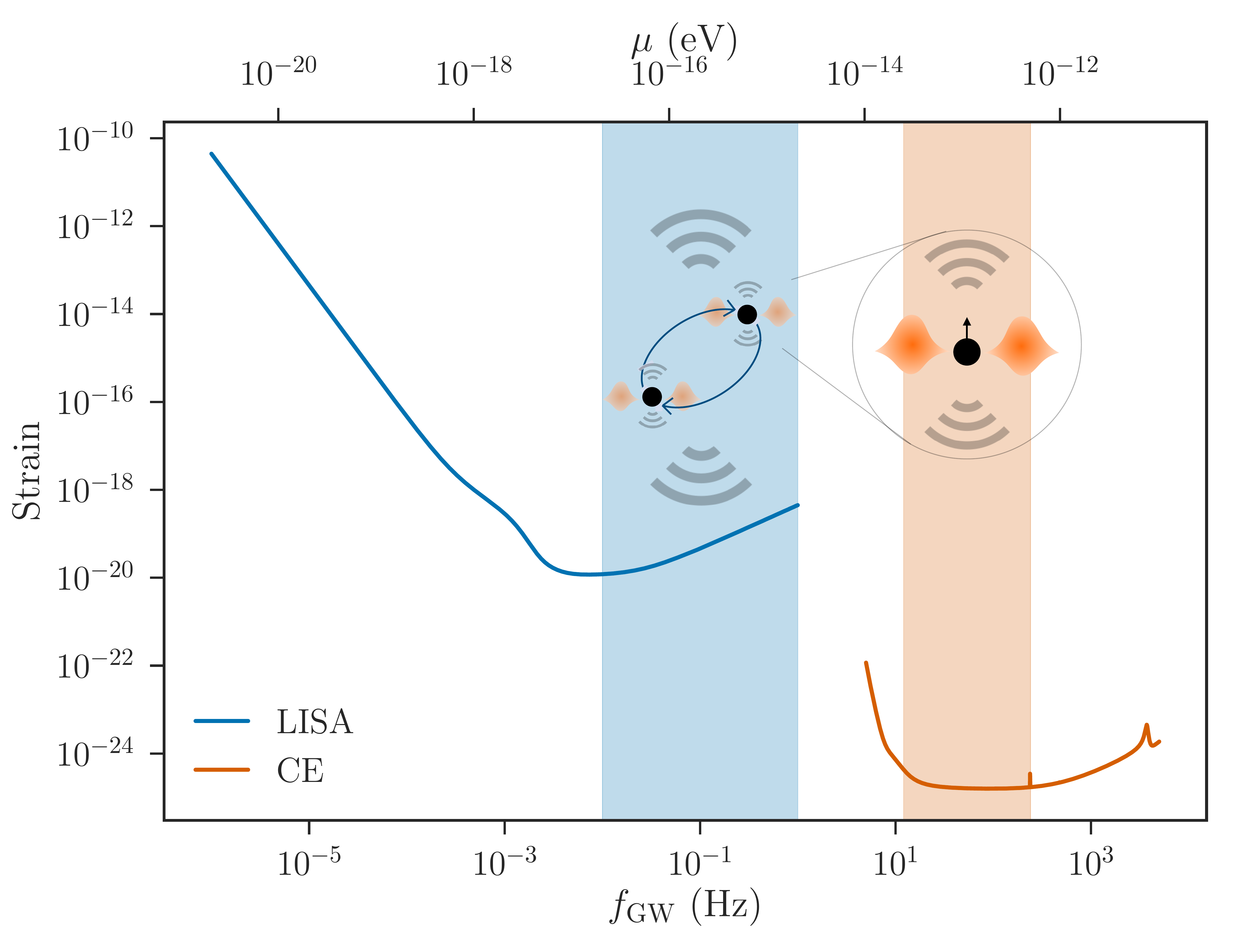}
\caption{
By analyzing \ac{BBH} inspirals with GW frequencies $0.01\lesssim f_{\rm GW} / \mathrm{Hz} \lesssim  1$ (shaded blue band), LISA will be able to provide crucial information for the search of GW signals from bosons with masses $25 \lesssim \mu / (10^{-15}\,\mathrm{eV}) \lesssim  500$ (shaded orange band) using \ac{CE}, or other 3G detectors, on the ground.
Solid curves mark the expected amplitude spectral densities of LISA (blue) and \ac{CE} (orange).
We explore the details of this application of multibanding starting in Sec.~\ref{sec:observation}, after providing some background in Sec.~\ref{sec:background}.
}
\label{fig:cartoon}
\end{figure}

The physical phenomenon at the core of this program is the proposed superradiant amplification of ultralight-boson fields around fast-spinning \acp{BH} \cite{Arvanitaki:2010sy,Yoshino:2013ofa,Yoshino:2014wwa,Arvanitaki:2014wva,Arvanitaki:2016qwi,Brito:2017wnc,Brito:2017zvb,Baryakhtar:2017ngi}.
Indeed, if a boson exists whose Compton wavelength is commensurate with the size of astrophysical \acp{BH}, its presence could be revealed by the spontaneous growth of a macroscopic, coherent quantum state in the \ac{BH} potential well --- a ``cloud,'' containing up to ${\sim}10\%$ of the mass of its  \ac{BH} host \cite{East:2017ovw, Herdeiro:2017phl, East:2018glu}.
After a short period of exponential growth, the cloud is expected to stabilize and emit quasi-monochromatic (``continuous'') GWs, potentially detectable by instruments on the ground or in space.
In the absence of boson self-interactions, this continuous GW signal may persist for a long time ($1-10^4$ yr for our parameters of interest) until the totality of the cloud has been radiated away \cite{Brito:2017zvb,Baryakhtar:2017ngi},

Detectors on the ground are most sensitive to GWs between ${\sim}[10,\,10^3]\,\mathrm{Hz}$, which makes them suitable probes of boson masses in the range $[10^{-14},\,10^{-12}]\,\mathrm{eV}$ \cite{Arvanitaki:2014wva,Brito:2017zvb}.
Although clouds formed by such bosons would not be directly detectable by LISA, the \ac{BH} harboring the cloud may itself lie in a binary~\cite{Zhang:2018kib,Baumann:2018vus,Berti:2019wnn,Baumann:2019ztm}; the binary would, in turn, emit GWs in the LISA band of ${\sim}[10^{-3},\,1]\,\mathrm{Hz}$ during its inspiral.
As explored below, this can happen for binaries with total mass in the range ${\sim}[\MtotLow,\, \MtotHigh]\,\Msun$, and bosons with masses within ${\sim}[\muLow,\, \muHigh]\times10^{-15}\,\mathrm{eV}$ (Fig.~\ref{fig:cartoon} and Table \ref{tab:params}).

Searches for continuous GWs are drastically simpler by knowledge of the source sky location and orientation, as well as estimates of the expected signal frequency and frequency derivative---all of which can be obtained from the inspiral signal.
This means that LISA will provide the information required for 3G detectors on the ground, like \ac{ET} and \ac{CE}, to conduct directed follow-up searches for ultralight-boson signals.
For simplicity, we will treat the case of LISA and a single \ac{CE} instrument on the ground, but our conclusions are easily generalizable to arbitrary detector networks.

We begin by reviewing some essential background on the dynamics of boson condensates and their associated GW signals in Sec.~\ref{sec:background}.
We then outline our proposed observation strategy in Sec.~\ref{sec:observation}, describing how a \ac{BBH} detection in LISA can inform a \ac{CW} follow-up by \ac{CE}.
In Sec.~\ref{sec:results}, we report our results on the feasibility of this measurement across the accessible parameter space, including the effect of tidal resonances, and discuss its interpretation in Sec.~\ref{sec:interpretation}.
We offer concluding remarks in Sec.~\ref{sec:conclusion}.

\begin{table}
\begin{ruledtabular}
	\caption{Relevant parameters (approximate ranges).}
	\label{tab:params}
\begin{tabular}{lcl}
		Boson rest mass & $\mu$ & $[\muLow,\, \muHigh]\times10^{-15} ~ \mathrm{eV}$\\
		CW frequency (CE) & $\fCW$ & $[\fCWLow,\, \fCWHigh] ~\mathrm{Hz}$\\
    \midrule
		Binary total mass & $\Mtot$ & $[\MtotLow,\, \MtotHigh] ~ \Msun$\\
		CBC frequency (LISA) & $f_{\rm LISA}$ & $[0.001,\, 1] ~\mathrm{Hz}$\\
	\end{tabular}
\end{ruledtabular}
\end{table}

\section{Background}
\label{sec:background}

In this section, we describe how an ultralight boson field can interact with a fast-spinning \ac{BH} to spontaneously give rise to a macroscopic boson cloud (Sec.~\ref{sec:formation}), which in turn proceeds to radiate CWs for a long time (Sec.~\ref{sec:emission}) if it is not disrupted by tidal resonances (Sec.~\ref{sec:resonances}).

\subsection{Cloud formation} 
\label{sec:formation}

Excited states of a boson field with mass $\bosonmass \equiv \mu/c^2$, and angular frequency $\omega \approx \omega_b \equiv \mu /\hbar$, will be superradiantly scattered off a Kerr \ac{BH} if \cite{Bekenstein:1973mi,Bekenstein:1998nt,Arvanitaki:2009fg,Brito:2014wla,Brito:2015oca}
\beq \label{eq:sr}
\ob / \mb < \obh\, ,
\eeq
where $\mb$ is the azimuthal quantum number of the boson's total angular momentum along the \ac{BH} spin direction, and $\obh$ is the angular frequency of the hole's exterior horizon (see, e.g., \cite{Teukolsky:2014vca}).
A boson state that satisfies this \emph{superradiant condition} will emerge with an enhanced amplitude (increased occupancy number) from interactions with the \ac{BH} \cite{Ternov:1978gq,Zouros:1979iw,Detweiler:1980uk,Dolan:2007mj,Arvanitaki:2009fg,Arvanitaki:2010sy,Brito:2014wla,Brito:2015oca,Brito:2017zvb,Witek:2012tr,East:2017mrj,Baryakhtar:2017ngi}, by extracting energy from its ergoregion in a manner fully analogous to the classical Penrose process \cite{Penrose:1969pc,zeldovich1,zeldovich2,Press:1972zz,Bekenstein:1973mi,Bekenstein:1998nt}.
(See \cite{Brito:2015oca} for a review on superradiance.)

A boson with Compton wavelength ($\lc \equiv h c /\mu$) comparable to the \ac{BH} size ($\rg \equiv G M /c$), may support semibound states in the \ac{BH} potential, with an energy-level structure analogous to the electronic levels in the hydrogen atom \cite{Ternov:1978gq,Zouros:1979iw,Detweiler:1980uk,Dolan:2007mj}.
The level spacing is controlled by a system-specific parameter $\alpha$ with the same role as the fine-structure constant in the hydrogen atom. 
This is given by the ratio of the two relevant length scales:
\beq \label{eq:alpha}
\fs \equiv \frac{\rg}{\lbc} = \frac{G \mbh}{c} \frac{m_b}{\hbar} = \frac{G \mbh}{c^3} \ob_b\, ,
\eeq
where $\lbc \equiv \lc / (2\pi)$ and $M$ is the \ac{BH} mass.
For superradiant energy levels to exist, \eq{sr} demands:
\beq \label{eq:sr2}
\fs < \frac{1}{2} \mb \chi  \left(1 + \sqrt{1-\chi^2}\right)^{-1} < \frac{m}{2}\, ,
\eeq
where $\chi$ is the \ac{BH}'s dimensionless spin, and the second inequality is obtained by noting $0\leq \chi < 1$.

If a superradiant state exists and boson self-interactions can be neglected, the occupancy number of the superradiant level will grow exponentially at the expense of the \ac{BH} \cite{Press:1972zz,Damour:1976kh,Zouros:1979iw,Detweiler:1980uk,Furuhashi:2004jk,Strafuss:2004qc,Dolan:2007mj}.
Superradiant growth can begin spontaneously, starting from quantum fluctuations in the field, and can continue to extract up to ${\sim}10\%$ of the \ac{BH} mass \cite{East:2017mrj,East:2017ovw,Herdeiro:2017phl}.
The process efficiently harvests angular momentum, greatly reducing the \ac{BH} spin, until superradiance shuts down when \eq{sr} is saturated.
As this happens, the final \ac{BH} spin $\chi_f$ asymptotically approaches
\beq \label{eq:final_spin}
\chi_f = \frac{4 \alpha_f \mb}{4\alpha_f^2 + \mb^2}\, ,
\eeq
with $\alpha_f$ computed for the {\em final} \ac{BH} mass.
If only one level is populated, then the final cloud mass will be \cite{Brito:2017zvb}
\beq \label{eq:mc}
\mc = \mbh_i - \mbh_f \approx M_i \frac{\alpha_i \chi_i}{\mb}\,,
\eeq
where $\mbh_i$ and $\mbh_f$ are the initial and final \ac{BH} masses, $\chi_i$ is the initial \ac{BH} spin, and the approximate equality holds for $\alpha_i\lesssim 0.1$.
In these approximations, the $\alpha$ computed for the {\em initial} \ac{BH} mass, $\alpha_i$ is larger than $\alpha_f$ by ${\sim}10\%$.
A more exact value for this quantity may be obtained by numerically solving a set of differential equations, e.g.~Eqs.~(17)--(21) in Ref.~\cite{Brito:2014wla}, assuming a quasiadiabatic evolution.
In this paper we follow this numerical approach, applying the same methods as in Ref.~\cite{Isi:2018pzk}.

Although a given system may support multiple superradiant levels, the different growth rates usually ensure that there's a single, fastest-growing state that is relevant at any given time.
For a field with spin-weight $s=0,1$ and a \ac{BH} with dimensionless angular frequency $\obhn \equiv \chi/ (2 + 2\sqrt{1-\chi^2})$, the level with the fastest superradiant growth will have angular quantum numbers $\{\jb,\, \lb,\, \mb\}$ such that
$
\jb = \lb + s = \mb = {\rm ceil}(\fs/\obhn)\,
$
\cite{Isi:2018pzk}, and the smallest possible radial quantum number that yields a boson energy satisfying \eq{sr}.
Consequently, the fastest-possible growing level over {\em all} values of $\fs$ and $\chi$ will be
$
\jb= \lb + s = \mb = 1~,~\nr=0\, .
$
In this paper, we will focus on scalar bosons ($s=0$), for which the overall dominant level has $\ell=m=1,\, n=0$.

The time it takes a single-level cloud to achieve its full size can be characterized by the $e$-folding time in the occupation number of the relevant quantum state, $\tinst$.
For the dominant scalar level, in the nonrelativistic limit ($\alpha_i \ll 1$), this is \cite{Brito:2017zvb}
\beq \label{eq:tinst_scalar}
\tinst^{\rm (s)} \approx 27\, {\rm days} \left(\frac{\mbh_i}{10\, \msun}\right) \left(\frac{0.1}{\alpha_i}\right)^9 \frac{1}{\chi_i-\chi_f}\, .
\eeq
Once superradiance has shut down and the cloud has reached its maximum size, the \ac{BH} and boson condensate become energetically decoupled.
The cloud may persist for a long time, until its energy is depleted through GW emission or resonant perturbations, as we review below.

\subsection{Dissipation due to GW emission}
\label{sec:emission}

The macroscopic, coherent quantum state that makes up the boson cloud can be thought of as a classical system, with a time-varying stress-energy tensor corresponding to the square-density of the field amplitude \cite{Arvanitaki:2010sy,Yoshino:2013ofa,Yoshino:2014wwa,Arvanitaki:2014wva,Arvanitaki:2016qwi,Brito:2017wnc,Brito:2017zvb,Baryakhtar:2017ngi}.
This time-varying stress-energy leads to the emission of GWs, which slowly carry away the energy contained in the cloud, as bosons annihilate into gravitons \cite{Arvanitaki:2010sy,Arvanitaki:2014wva}.
For the dominant scalar level, it may be shown that this results in the emission of a quasimonochromatic \ac{CW}, with initial frequency approximated by  ($\alpha_i \ll 1$)
\beq \label{eq:fgw_approx}
\fgw \approx \frac{\mu}{\hbar\pi} \approx 645\, {\rm Hz} \left(\frac{10 \msun}{\mbh}\right) \left(\frac{\alpha_i}{0.1}\right) ,
\eeq
and corresponding initial (angle-averaged) strain amplitude 
\begin{align} \label{eq:h0_approx}
h_0 \approx 8\times10^{-28}\left( \frac{\mbh_i}{10\Msun} \right) \left( \frac{\alpha_i}{0.1} \right)^7 \left( \frac{\text{Mpc}}{\DL} \right) \left(\frac{\chi_i-\chi_f}{0.1}\right),
\end{align}
where \DL is the cloud's luminosity distance \cite{Yoshino:2014wwa,Brito:2014wla,Arvanitaki:2016qwi}.
For $\alpha_i \gtrsim 0.1$, this approximation breaks down and Eq.~\eqref{eq:h0_approx} tends to overestimate the GW power \cite{Brito:2017zvb}.
Instead, we estimate $h_0$ more accurately using the numerical results from \cite{Brito:2017wnc,Brito:2017zvb}.
We define the strain amplitude for the quadrupolar mode following the usual LIGO-Virgo convention for the \ac{CW} plus and cross polarizations,%
\footnote{We factor in the correction in the Erratum of \cite{Isi:2018pzk}.}
\beq \label{eq:plus}
h_+ = \frac{1}{2}h_0 \left(1+\cos^2\iota\right) \cos[\Phi(t)]\, ,
\eeq
\beq \label{eq:cross}
h_{\times} = h_0 \cos\iota \sin[\Phi(t)]\, ,
\eeq
where $\iota$ is the source inclination (angle between the \ac{BH} spin axis and the line of sight), and $\Phi(t)$ is the sinusoidal phase evolution implied by $f$.
For a BBH with aligned spins, i.e. spins parallel to the orbital angular momentum, the cloud's inclination $\iota$ is the same as the inclination $\iota_{\rm orb}$ of the binary's orbital plane.
In general, $\iota$ and $\iota_{\rm orb}$ need not be equal for a BBH with precessing spins.

Since GW emission removes energy-momentum from the cloud, the cloud mass gradually decreases, causing the signal frequency and amplitude to evolve \cite{Baumann:2018vus}.
Initially, the GW frequency increases with an approximate time derivative (with respect to the proper time)
\begin{equation} \label{eq:fdot}
\dot{f} \approx 3\times 10^{-14} \, \text{Hz}/\text{s} \left( \frac{10\Msun}{M_i} \right)^2 \left( \frac{\fs_i}{0.1} \right)^{19} (\chiIni-\chi_f)^2 \, ,
\end{equation}
for $\alpha_i \ll 1$ \cite{Isi:2018pzk,Baryakhtar:2017ngi}.
This spin up is characteristic of gravitationally bound systems, and distinguishes boson clouds from other potential \acp{CW} sources, like nonaxisymmetric neutron stars (see \cite{Riles:2017evm} for a review).
The frequency derivative itself evolves slowly, following $\dot{f} \propto - \dot{M}_c(t) \propto \left( 1+t/\tauGW \right)^{-2}$, where $\tauGW$ is a characteristic timescale~\cite{Isi:2018pzk,Baryakhtar:2017ngi},
\begin{align}\label{eq:tauGW_approx}
\tauGW \approx 6.5\times 10^4 \, \mathrm{yr} \left( \frac{\mIni}{10\Msun} \right) \left( \frac{0.1}{\alpha_i} \right)^{15} \frac{1}{\chiIni-\chi_f} \, .
\end{align}
We incorporate this trend approximately by writing
\begin{equation}\label{eq:f_evol}
f(t) \approx f_0 + \dot{f}_0\tauGW\left(1-\frac{1}{1+t/\tauGW}\right),
\end{equation}
where $f_0$ and $\dot{f}_0$ are the initial frequency and frequency derivative respectively.
Besides affecting the frequency, the reduction in the boson cloud mass causes the GW amplitude to decay~\cite{Baumann:2018vus},
\begin{align} \label{eq:h0_evolution}
h(t) \propto h_0 \left(1+t/\tauGW\right)^{-1},
\end{align}
with the initial $h_0$ determined numerically.

\subsection{Depletion due to orbital resonances}
\label{sec:resonances}

The tidal perturbation of a massive companion in a corotating (counterrotating) orbit may introduce hyperfine (Bohr) resonances in the boson energy levels.
This occurs when the orbital frequency matches the energy gap between superradiant and decaying modes (that is, modes with a positive imaginary part of their eigenfrequencies)~\cite{Zhang:2018kib,Baumann:2018vus,Berti:2019wnn,Baumann:2019ztm}.
If this occurs, then the boson cloud depletes through the excitation of a decaying mode.
The characteristic strength and timescale of the depletion depends on the ratio between the mass of the perturber and the \ac{BH} hosting the cloud.
As a conservative estimate, we will assume that the cloud becomes totally depleted once the orbit hits a resonance frequency.
This is generally valid for equal mass \ac{BBH} systems, for which the decay rate of the resonant mode is high and the decay timescale is much shorter than the orbital timescale~\cite{Zhang:2018kib,Baumann:2018vus,Berti:2019wnn,Baumann:2019ztm}.
To a good approximation, the observable result of this resonance will be to instantaneously turn off the GW signal.

For a \ac{BBH} system in which the components have similar masses but opposite spins, the frequency associated with the hyperfine resonance will always be lower than of the Bohr resonance~\cite{Baumann:2018vus,Berti:2019wnn}.
This means that, as a binary evolves from larger to smaller separations, the former will be the first to become relevant.
Hence, we restrict our attention to hyperfine resonances, the inspiral GW frequency of which is given by
\begin{align}\label{eq:fres}
	\fres = \frac{1}{12\pi}\mu\chi_{f}\alpha_{f}^5\, ,
\end{align}
where $\chi_f$ is the \ac{BH} spin at the saturation of superradiance given by \eq{final_spin}.
For $\alpha_i \ll 1$, $\chi_f \approx 4\alpha_i$, and
\begin{align}\label{eq:fres_approx}
	\fres \approx \frac{1}{3\pi M_i} \alpha_i^7\, .
\end{align}

\begin{figure}[tb]
\centering
\includegraphics[width=0.9\columnwidth]{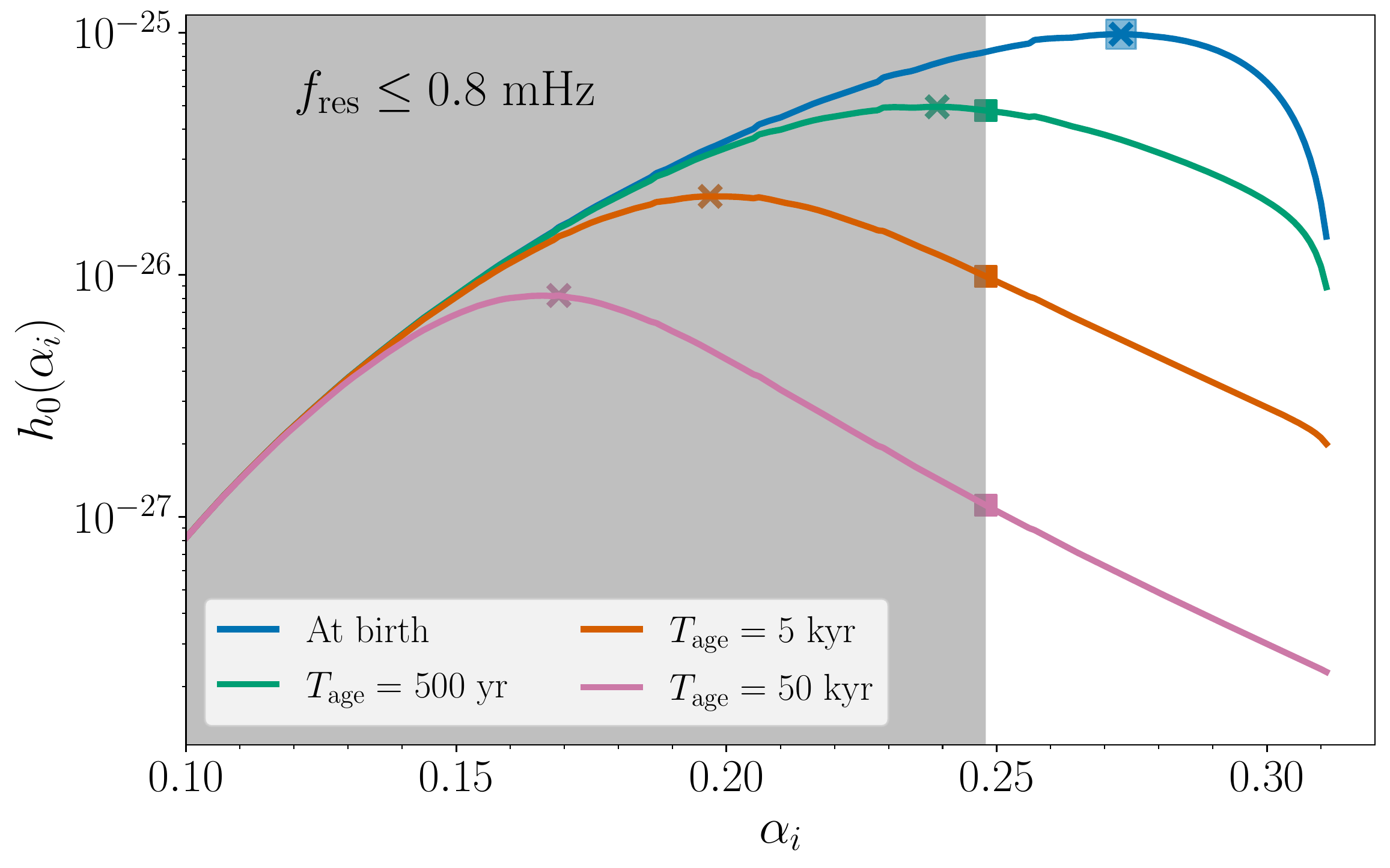}
\caption{
Strain amplitude as a function of $\alpha_i$ for $M_i=1000\,\Msun$, $\chiIni=0.9$, and $\DL=400$~Mpc.
The grey area indicates the $\alpha_i$'s that are not observable by our technique due to resonant depletion.
Solid lines with different colors correspond to various $\Tage$: zero age (blue), 500~yr (green), 5~kyr (orange), and 50~kyr (pink).
On each colored line, the crossmarker denotes the optimal $\alpha_i$ which generates the maximum amplitude without resonance, while the squaremarker corresponds the best $\alpha_i$ conditioned by resonance.
Two maxima are the same only for zero age.
The optimal $\alpha_i$'s before resonance are not observable for other $\Tage$.
}
\label{fig:h_alpha}
\end{figure}

Because of resonances, the presence of a binary companion can restrict the \ac{CW} power that may be expected from a boson cloud around a given \ac{BH}:
the boson mass that would produce the strongest signal for an isolated \ac{BH} may also lead to resonances that destroy the cloud if the \ac{BH} is in a binary.
We illustrate this in Fig.~\ref{fig:h_alpha} for a \ac{BH} with $M_i=1000\,\Msun$, $\chiIni=0.9$, and $\DL=400$~Mpc.
The different curves represent the \ac{CW} amplitude $h_0$, as in Eqs.~\eqref{eq:plus}~and~\eqref{eq:cross}, for different boson masses (parametrized by $\alpha_i$), with color representing the cloud age $\Tage$, i.e.~the time after the end of superradiance.
For a given $\Tage$ the emission peaks for the value of $\alpha_i$ that corresponds to a boson optimally matching the \ac{BH}, which we indicate by a cross.
Higher $\alpha$'s lead to faster depletion by Eq.~\eqref{eq:tauGW_approx}, so the crosses move left as \Tage increases in Fig.~\ref{fig:h_alpha}.
If we rely on LISA to identify potential \ac{CW} sources, then the cloud needs to persist at least until the binary enters the LISA frequency band.
As we will see below, this means that the resonances should not occur at frequencies $\fres < 0.8\, \mathrm{Hz}$, or the system will not be detected (gray region in Fig.~\ref{fig:h_alpha}).
Depending on the cloud age, this means that the loudest \ac{CW} we expect to see from a \ac{BH} in a binary is generally weaker than it would have been if the \ac{BH} had been isolated (marked by boxes in Fig.~\ref{fig:h_alpha}).
For our example system, this is the case for all shown $\Tage$'s except $\Tage=0$ (blue curve).

Cloud resonances may backreact on the binary's orbit, inducing a dephasing on inspiral GW waveforms~\cite{Zhang:2018kib,Baumann:2019ztm}.
Looking for such kind of dephasing may be another smoking-gun evidence of the existence of boson.
However, as we will discuss below, our observation strategy requires a sufficiently long segment of inspiral in LISA and \ac{CW} in CE before the resonance occurs.
Therefore, we ignore the effect of backreaction on either the inspiral waveform or the cloud emission.

Finally, the above calculations for resonant depletion assume a perturbation of a weak tidal field generated by the companion object.
When the orbital separation reaches the Roche radius, the strong tidal perturbation may also disrupt the boson cloud significantly~\cite{Cardoso:2020hca}.
According to Ref.~\cite{Cardoso:2020hca}, the critical frequency $f_{\mathrm{crit}}$ of this tidal disruption for an equal-mass binary is
\begin{align}
	f_{\mathrm{crit}} \approx \frac{1}{\sqrt{250}\pi M_i}\alpha_i^3\, .
\end{align}
Comparing to Eq.~\eqref{eq:fres_approx}, the ratio of this critical frequency to the resonance frequency is
\begin{align}
	\frac{f_{\mathrm{crit}}}{\fres} \approx 0.2\alpha_i^{-4}\, .
\end{align}
Since the $\alpha_i$'s we will be interested lie in the range $\sim\left[0.15,0.45\right]$ (Sec.~\ref{sec:results:ideal}), we expect the boson clouds we target to deplete due to orbital resonances before having a chance of being tidally disrupted.

\section{Observation strategy/scenario}
\label{sec:observation}

A \ac{BBH} signal detected by LISA can provide crucial information about the location and properties of the component \acp{BH}, allowing detectors on the ground to conduct a directed follow-up for ultralight bosons.
For this to be possible, the binary must do the following:
\begin{enumerate}[label={(\roman*)}]
\item have a total mass and initial orbital separation such that the inspiral signal remains in the LISA frequency band sufficiently long to be detectable; 
\item have component \acp{BH} in the stellar-mass range, so as to potentially host boson clouds radiating GWs in the \ac{CE} (or \ac{ET}) frequency band.
\end{enumerate}
These conditions can be satisfied by equal-mass binaries with total mass $\Mtot$ in the range ${\sim}\left[\MtotLow,\, \MtotHigh\right]\Msun$.
Assuming a \ac{SNR} threshold of 8, LISA can detect such systems over a 4 yr observation period if the initial orbital separation yields a starting GW frequency of $\fIniLISA \sim 0.01\, \mathrm{Hz}$.
Meanwhile, component \acp{BH} with masses in the range ${\sim}\left[\McompLow,\, \McompHigh\right]\Msun$ may host boson clouds sourcing GWs with frequencies in the range ${\sim}\left[\fCWLow,\, \fCWHigh \right]\mathrm{Hz}$ [cf.~Eq.~\eqref{eq:fgw_approx}], well within the most-sensitive band of future ground-based detectors.
Assuming $\alpha_i \approx 0.1$, this corresponds to bosons with rest masses $\mu$ within ${\sim}\left[\muLow,\, \muHigh\right]\times10^{-15}\, \mathrm{eV}$ [cf.~Eq.~\eqref{eq:alpha}].
(See Fig.~\ref{fig:cartoon} and Table \ref{tab:params}.)

As the binary inspirals towards merger, the orbit may reach the resonance frequency of the cloud and destroy it (Sec.~\ref{sec:resonances}).
If this takes place before the BBH signal has entered the LISA frequency band, the cloud will have depleted before \ac{CE} has had a chance to observe it.
Therefore, for our strategy to be successful, the pair of $M_f$ and $\mu$ must lead to resonances that only take place \textit{after} the binary has entered the LISA band. 
Even then, resonances may hinder the \ac{CE} follow-up: if the resonance occurs \textit{during} the LISA observation, the cloud \ac{CW} will be terminated sooner and \ac{CE}'s available observation period will be reduced.
Below, we will factor this effect by computing the reductions in \ac{CE} \ac{SNR} expected from the resonances.

If at least one of the \acp{BH} in the binary indeed harbors a boson cloud actively emitting \acp{CW}, its detectability by \ac{CE} will further depend on the source sky location, orientation and distance from Earth, as well as the \ac{BBH} orbital parameters.
All of these properties can be inferred based on the LISA signal.
Below, we describe how the two measurements, by LISA and CE, would take place.

\subsection{LISA measurement}
\label{sec:lisa}

LISA will be able to detect the inspiral stage of compact binaries with total mass above $10\,\Msun$.
Such signals will carry information about the masses and spins of the component \ac{BH}, as well as the system's sky location, luminosity distance and orbital parameters (including semimajor axis, orbital phase, and eccentricity).
How well these properties can be extracted from the data will depend, in part, on the \ac{SNR} of each particular signal.

To calculate inspiral \ac{SNR}s in LISA, we follow Ref.~\cite{Cornish:2018dyw} and assume the full, 2.5 million-km-long, three-arm configuration as proposed for the ESA L3 mission~\cite{LISA}.
For a face-on binary, i.e., one whose orbital plane is perpendicular to the line of sight, the optimal sky-averaged \ac{SNR} of LISA is given by
\begin{align}
	\rhoLISA = 8 \int_{\fIniLISA}^{f(\TLISA)}\frac{A^2(f)}{\SLISA(f)} df,
\end{align}
where $\fIniLISA$ is the initial GW frequency, $\TLISA$ is the observation duration, $A(f)$ is the amplitude of the inspiral waveform, and $\SLISA(f)$ is the instrument's noise \ac{PSD}.
We consider an inspiral signal to be detectable if $\rhoLISA\geq8$.
We assume the sky-location-averaged sensitivity from the analytical fit in Ref.~\cite{Cornish:2018dyw} (Fig.~\ref{fig:cartoon}), which accounts for both instrumental and galactic confusion noise, and assume an observation time $\TLISA=4$~years.

Since the binaries we are interested in only inspiral in the LISA band, merging at much higher frequencies, we use the stationary phase approximation for a circular, nonspinning quadruple system to write~\cite{Sathyaprakash:2009xs}
\begin{align}
	A(f) = \sqrt{\frac{5}{24}}\frac{\mathcal{M}f^{-7/6}}{\pi^{2/3}D_L},
\end{align}
where $\mathcal{M}=(1+z)(M_1M_2)^{3/5}/M_{\rm tot}^{1/5}$ is the redshifted chirp mass and $D_L$ is the luminosity distance.
For a given $\TLISA$ and $\fIniLISA$, we evolve the system to obtain the final frequency $f(\TLISA)$, using Peters' formula~\cite{Peters:1964}.
Since $\fIniLISA$ only depends on the initial binary separation, LISA may observe binaries with any $\fIniLISA$ within its sensitive frequency band.
However, to select systems with minimal loss due to resonant depletion, we demand $\fIniLISA$ to be the lowest possible, so as to get $\rhoLISA=8$. 
Hence $\fIniLISA$ is different for different $\Mtot$, e.g. $\fIniLISA \approx 1$~mHz for $\Mtot \approx 1000\Msun$ but $\fIniLISA\approx 6$~mHz for $\Mtot \approx 100\Msun$, assuming $D_L = 400$ Mpc.

To estimate the impact of LISA measurement uncertainty in the observable range of boson mass, we take the typical $1\sigma$ component mass and distance uncertainties to be ${\sim}10\%$ and ${\sim}20\%$, from Refs.~\cite{Sesana:2016ljz} and~\cite{DelPozzo:2017kme}, respectively.
For concreteness, below we restrict our attention to nearly equal mass binaries.

If, in addition to the \ac{BBH} masses, LISA could accurately determine the component spins $\chi_{1,2}$, Eq.~\eqref{eq:final_spin} would immediately exclude boson masses with $\chi_f(\mu) \leq \chi_{1,2}$.
This would further narrow down \ac{CE}'s potential search space.
While the spin magnitude measurement can be as good as $\Delta \chi \approx 0.1$ for massive binaries of $\geq 10^4\Msun$~\cite{Klein:2015hvg}, there is a lack of detailed studies on the spin measurement for intermediate mass binaries.
In the following sections, we do not consider the constraint from individual spin measurements.

\subsection{Ground-based follow-up}

The strategy for \ac{CE} follow-up is similar to that of boson clouds forming around \ac{CBC} remnants \cite{Isi:2018pzk,Ghosh:2018gaw}: information about the source location from LISA vastly simplifies what would otherwise be an extremely expensive all-sky search for \acp{CW} from boson clouds, and information about the \ac{BH} mass significantly narrows down the expected frequency space, through Eq.~\eqref{eq:fgw_approx}.
Unlike in the case of a \ac{CBC} remnant, however, LISA would provide us with post-, not pre-, superradiance \ac{BH} parameters (assuming a cloud is present).
This follows from the fact that, rather than newly born boson clouds, we should expect LISA to detect systems long after the cloud has formed.

Following up a \ac{BH} in a binary involves some additional complications with respect to solitary \acp{BH}.
Most prominently, the orbital motion will be imprinted in the boson \ac{CW} through a time-varying Doppler shift (R{\o}mer delay).%
\footnote{There will also be relativisitc effects, like the Shapiro delay, but those are not relevant for semicoherent \ac{CW} searches~\cite{Suvorova2016,Sun2018}.}
In the frequency domain, this has the effect of spreading the signal power over orbital sidebands centered around the intrinsic signal frequency $\fCW$.
An orbit with period $P$, semimajor axis $a$ and inclination $\iota$ will spread the GW power over a bandwidth (see Sec.~IIIE in Ref.~\cite{Sammut:2013oba})
\begin{align}\label{eq:orbitalB}
B	&\approx \frac{4\pi a \vert \sin{\iota} \vert \fCW}{cP} \nn \\
	&= 2 \vert \sin{\iota} \vert\fCW \left( \frac{G \Mtot}{c^3} \fLISA \right)^{1/3},
\end{align}
where we have assumed the orbit is Keplerian to write $B$ in terms of $\Mtot$ and $\fLISA$, the GW inspiral frequency that would be seen by LISA.
For example, for an equal-mass binary with $\Mtot=1000\,\Msun$ inspiraling at $\fLISA\sim1$~mHz, and a boson cloud radiating at $\fCW=40$\,Hz, the bandwidth is roughly $B \approx 1.3 \vert \sin{\iota} \vert$~Hz.

There exist several established methods for detecting \acp{CW} from sources in binaries with known sky locations \cite{Messenger:2007zi, Sammut:2013oba, Whelan:2015dha, Suvorova2016, Suvorova:2017dpm}, all of which can collect the power distributed across the orbital sidebands.
In principle, knowledge of the orbital parameters (including the phase) can be used to fully demodulate the signal and recover all signal power, so that the search sensitivity is not impacted by the orbital motion.
In our scenario, this will be made possible by the precise characterization of the orbit through the LISA inspiral measurement.
We will thus base our \ac{CE} sensitivity estimates on previous studies for boson signals from isolated sources in Ref.~\cite{Isi:2018pzk}.

The projections of \cite{Isi:2018pzk} were obtained through the search method described in Refs.~\cite{Suvorova2016,Suvorova:2017dpm,Sun2018}, although optimized for signals with a positive frequency derivative to accommodate Eq.~\eqref{eq:fdot}.
When targeting binary systems, uncertainties in the orbital parameters may be marginalized over with minimal impact on the sensitivity \cite{Suvorova:2017dpm}.
The measurement of $\iota$ is correlated with $\DL$ since both parameters affect the overall amplitude of the inspiral signal, and a recent case study suggests a full Bayesian parameter estimate may return a \textit{poor} $\iota$ measurement ${\sim}1$~rad for SNR $\sim 10$~\cite{Marsat:2020rtl}.
However, the usual implementation of this particular method parallelizes the search by splitting the frequency band in a way that requires $B \lesssim 0.5$~Hz, which would limit accessible inclinations (e.g.~$\iota \lesssim 10^{\circ}$ for the $1000\,\Msun$ binary considered above).
This limitation can be circumvented by reducing parallelization, at the expense of increased computing cost.

Although the search for boson signals in \ac{CE} data is conceptually no different from other directed \ac{CW} searches, one feature sets our scenario apart: depending on the mass ratio, we could expect \acp{CW} from clouds around \textit{both} binary components.
In fact, if the two \acp{BH} have near-equal masses and similar histories, they should both be compatible with the same set of boson masses and, thus, lead to \acp{CW} signals with the same intrinsic frequency given by Eq.~\eqref{eq:fgw_approx}.
The amplitudes of the two signals would depend on the individual \ac{BH} masses and histories (i.e.~cloud age and presuperradiance spin), following Eqs.~\eqref{eq:h0_approx} and \eqref{eq:h0_evolution}.
Assuming that the two clouds are formed simultaneously, the amplitude ratio of the two \ac{CW} signals $h_0^{(2)}/h_0^{(1)}$ has a steep dependence on the binary mass ratio,
\beq
h_0^{(2)}/h_0^{(1)} \approx (M_2/M_1)^8 \, ,
\eeq
for component masses $M_{1,2}$.
With a small asymmetry in BH masses $M_2/M_1 = 0.9$, the amplitude ratio drops to ${\sim} 40\%$.
Therefore, we generally expect one of the putative \ac{CW} signals to dominate.

Having two overlapping \ac{CW} signals could, at best, enhance the effective \ac{SNR} available to the search.
Assuming the power is added incoherently (as would be the case for the methods mentioned above), the improvement could be up to a factor of $\sqrt{2}$ for $M_1=M_2$.
Methods could be developed in the future to coherently track both signals, taking into account orbital phase, and thus achieve a factor of 2 improvement instead.
In any case, an amplitude enhancement is degenerate with a reduction in the luminosity distance, so all quantities of interest (like detection horizons) can be scaled trivially.
Therefore, in the detectability discussions below, we will simply assume a single signal is present at a time.

Below, we assume a single \ac{CE} instrument operating at the design sensitivity projected in~\cite{Evans:2016mbw,iswp2016}, and shown in Fig.~\ref{fig:cartoon}.
While \ac{CE} may observe boson \acp{CW} within $D_L\sim10$~Gpc~\cite{Isi:2018pzk} for the \ac{BH} masses of interest, LISA may detect binaries within $D_L\sim400$~Mpc $({\sim}1000\,\rm Gpc)$ for $\Mtot = \MtotLow\, \Msun (\MtotHigh\, \Msun)$, respectively~\cite{Jani:2019ffg}.
Therefore, we expect our observation technique to be limited by LISA (\ac{CE}) in the lower (upper) mass end, respectively.

\begin{figure*}
\centering
	\subfloat[$\mIni=50\Msun$ \label{fig:age_50Msun}]{\includegraphics[width=0.45\textwidth]{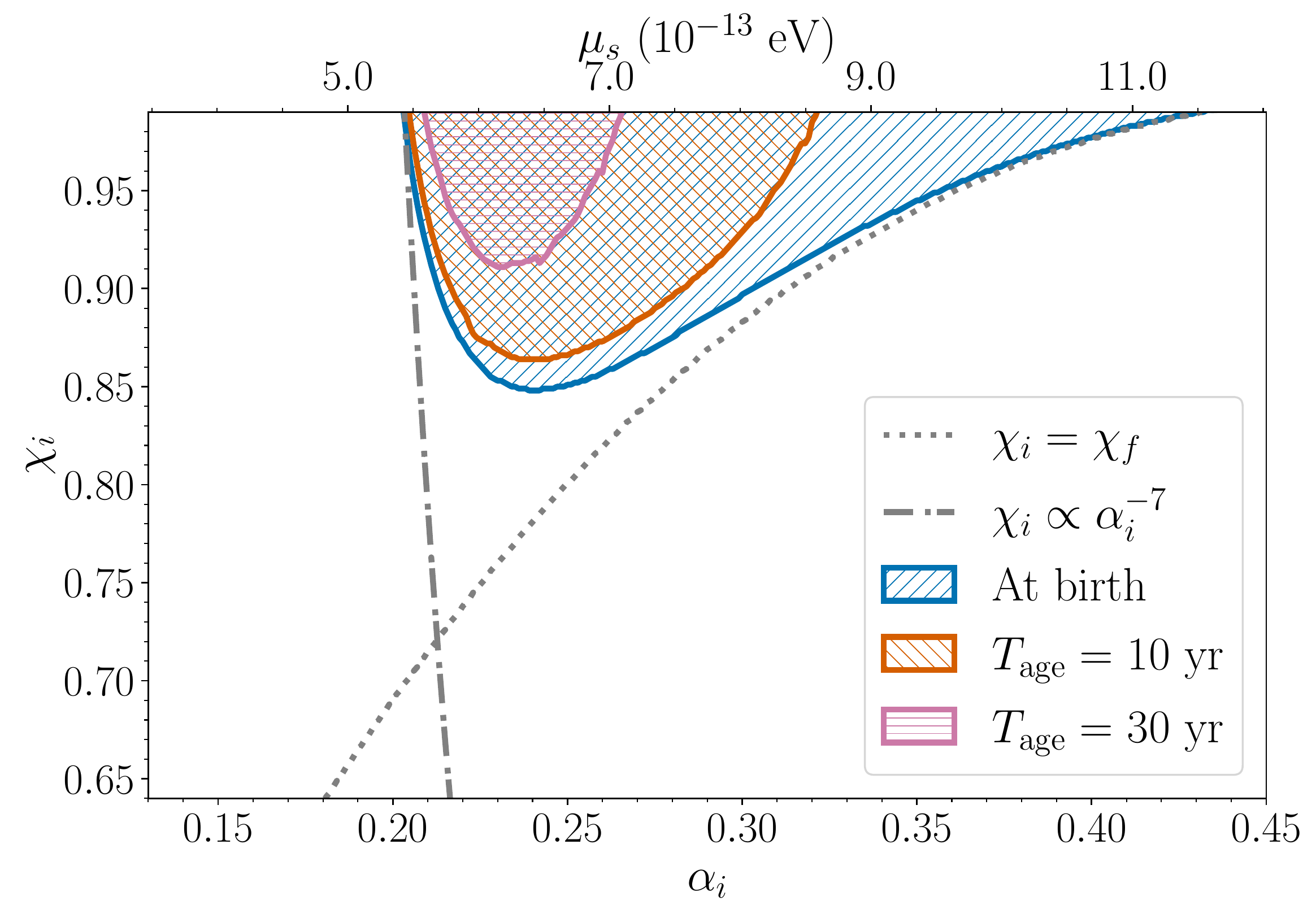}}
	\subfloat[\label{fig:age_1000Msun} $\mIni=1000\Msun$]{\includegraphics[width=0.45\textwidth]{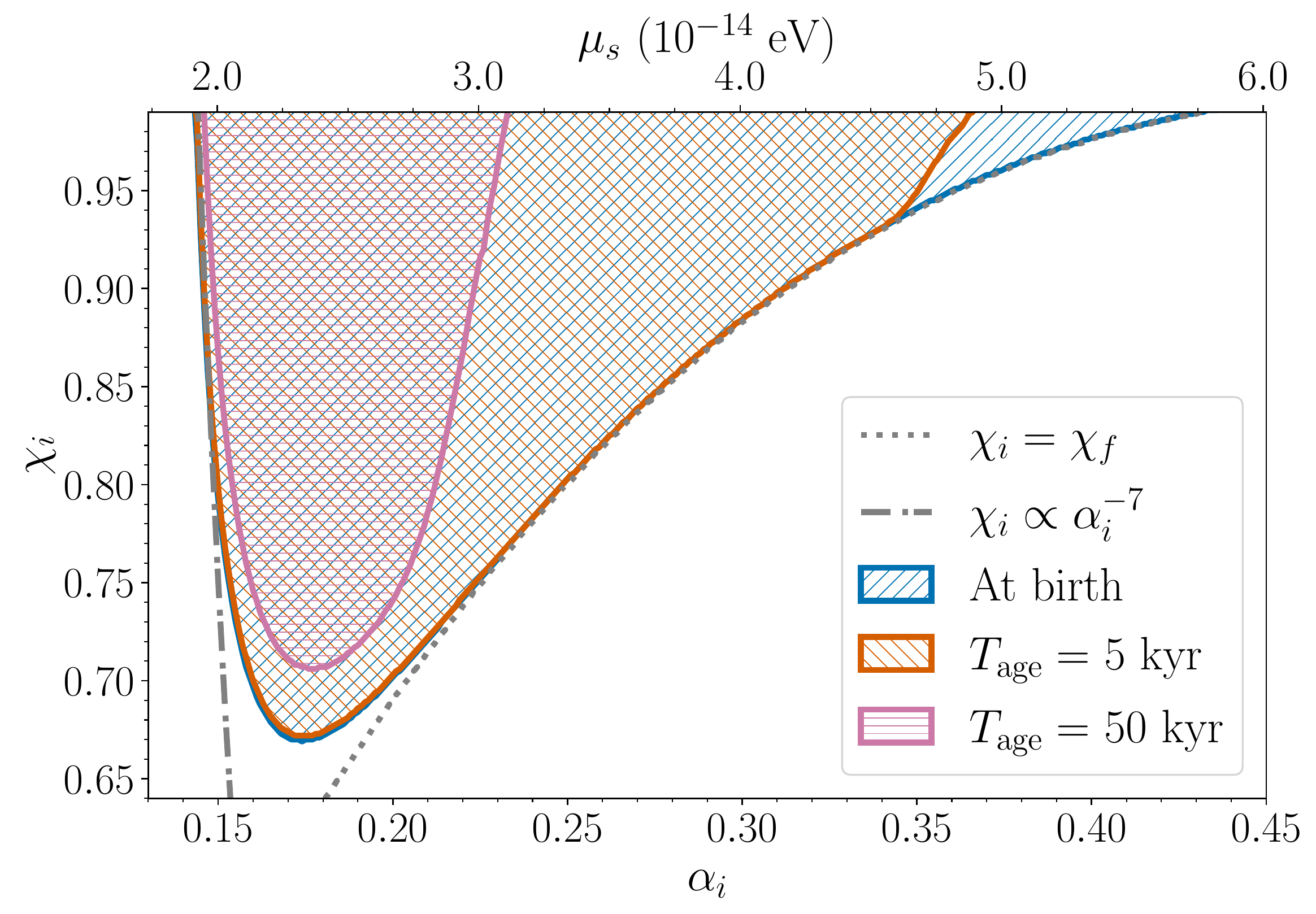}}
\caption{
Detectable parameter space $(\alpha_i,\chiIni)$ at $\DL=400$~Mpc for two example of \mIni's.
Each hatched region corresponds to the detectable space for a host \ac{BH} at different \Tage's: 10~yr (orange) and 30~yr (pink) for $50\,\Msun$; 5~kyr (orange) and 50~kyr (pink) for $1000\,\Msun$.
Older clouds have had time to dissipated more energy momentum, and (since the strain amplitude is proportional to the cloud mass) the detectable region decreases as \Tage increases.
The grey dotted line shows $\chi_f$ as a function of $\alpha_i$, which is the limit below which \eq{sr} is violated and superradiance cannot occur.
The grey dashed-dotted line shows the $\chiIni \propto \alpha_i^{-7}$ scaling in Eq.~\eqref{eq:h0_approx}, which suppresses the low $\alpha_i$ amplitude due to insufficient cloud mass.
In this and all other figures, we assume a \ac{CW} observation time of $T{\rm obs}=1$~yr.
}
\label{fig:exc_age}
\end{figure*}

\section{Detectable parameter space}
\label{sec:results}

Assuming LISA has detected a given \ac{BBH}, we would now like to quantitatively characterize the conditions needed for \ac{CE} to detect a \ac{CW} from a putative cloud in the system, and identify the boson masses that can be probed by such a measurement.
As mentioned above, the first requirement is that the total mass of the system be in the range
\beq
\MtotLow \lesssim \Mtot/\Msun \lesssim \MtotHigh\, .
\eeq
This is however not a sufficient condition: \ac{BH} age, presuperradiance spin, and tidal resonances may all limit the expected \ac{CW} amplitude and, thus, its detectability.
In this section, we explore how these factors affect potential boson-mass constraints (Sec.~\ref{sec:results:ideal}); we discuss the effect of uncertainties in the LISA measurement (Sec.~\ref{sec:results:uncertain}); and, finally, present detection horizons over parameter space (Sec.~\ref{sec:results:limits}).

Throughout, we consider that \ac{CE} is able to detect a given \ac{CW} if its amplitude exceeds a threshold $\hthr$, estimated using the scaling provided in Eq.~(38) of Ref.~\cite{Isi:2018pzk} for one detector, namely
\beq \label{eq:hthr}
\hthr(f) = 1.7 \times 10^{-26}
\left[\frac{S_n(f)}{S_{\rm ref}}\right]^\frac{1}{2}
\left[\frac{8\, {\rm d}}{T_{\rm drift}}\right]^\frac{1}{4}
\left[\frac{1\, {\rm yr}}{T_{\rm obs}} \right]^\frac{1}{4}
\eeq
where $S_n(f)$ is the noise \ac{PSD}, $S_{\rm ref}=1.6\times10^{-47}\,{\rm Hz}^{-1}$ is a reference value, $T_{\rm drfit}$ is the period of coherent segments of the \ac{CW}, and $T_{\rm obs}$ is the total observation time.
For the case of multiple detectors, $S_n(f)$ should be replaced by the harmonic mean of the corresponding \acp{PSD}.
As in Ref.~\cite{Isi:2018pzk}, we rescale $T_{\rm drift}=(2\dot{f})^{-1/2}$ to be the largest value allowable by the expected frequency evolution of Eq.~\eqref{eq:fdot} for a given boson mass.
We also take into account that the cosmological redshift scales down the detector frame $\dot{f}$ by $1/(1+z)^2$, as well as $f$ by $1/(1+z)$.
Throughout this study, we choose $T_{\rm obs}=1$~yr.

In order to estimate the \ac{CW} amplitude expected from a given system, we approximate the presuperradiance \ac{BH} mass by its postsuperradiance value, as would be provided by LISA.
This is equivalent to assuming a ${\sim}10\%$ error on the mass, which is comparable to the projected uncertainty in LISA's component mass measurement (see Sec.~\ref{sec:lisa}).%
\footnote{If needed, we could always (albeit at significant computational expense) remove this approximation by recursively solving the cloud evolution equations to find the pre-superradiance parameters that would yield a final \ac{BH} mass compatible with the LISA measurement.}

\begin{figure*}[tb]
\centering
	\subfloat[$\mIni=50\Msun$.]{\includegraphics[width=0.45\textwidth]{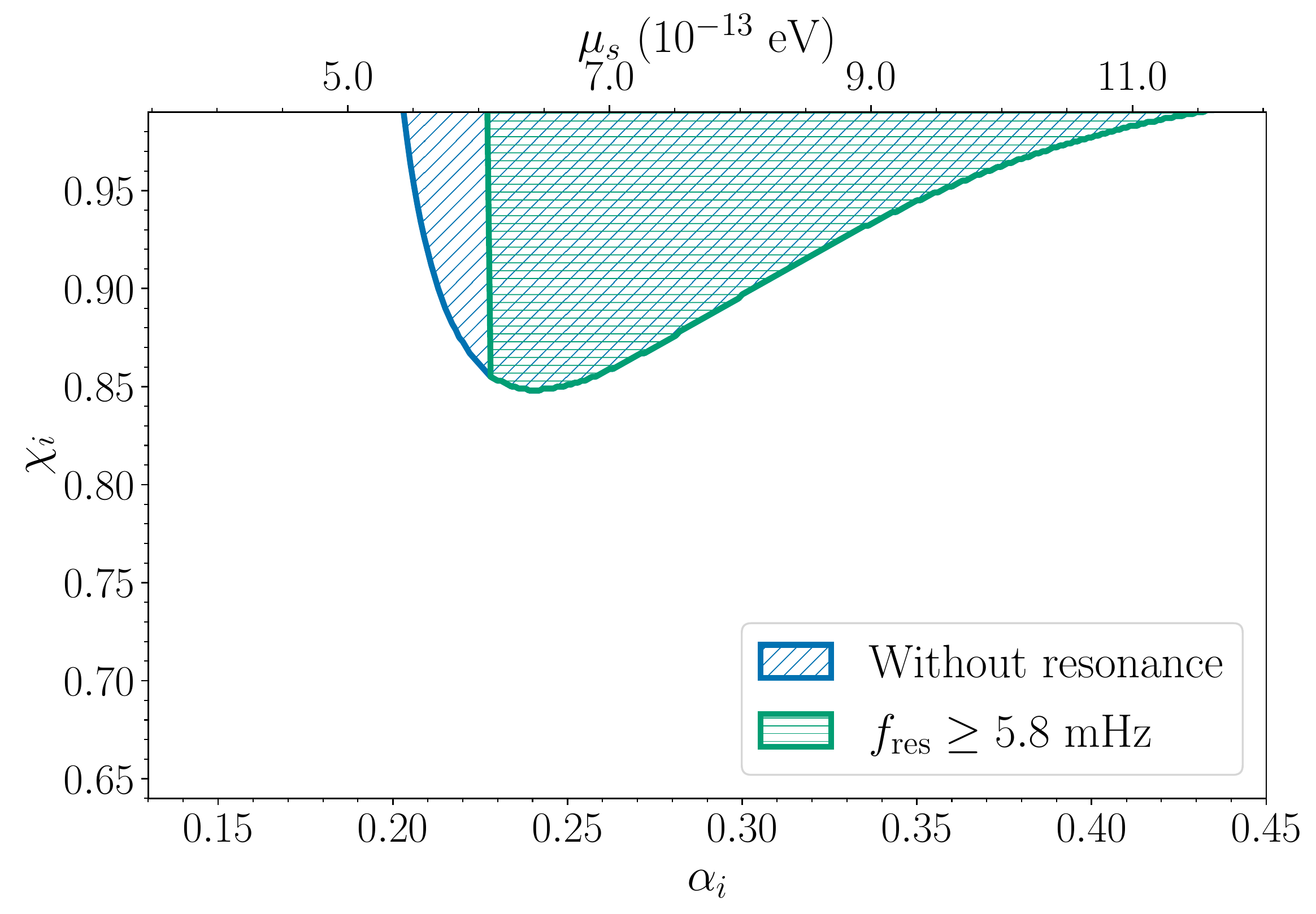}}
	\quad
	\subfloat[$\mIni=1000\Msun$.]{\includegraphics[width=0.45\textwidth]{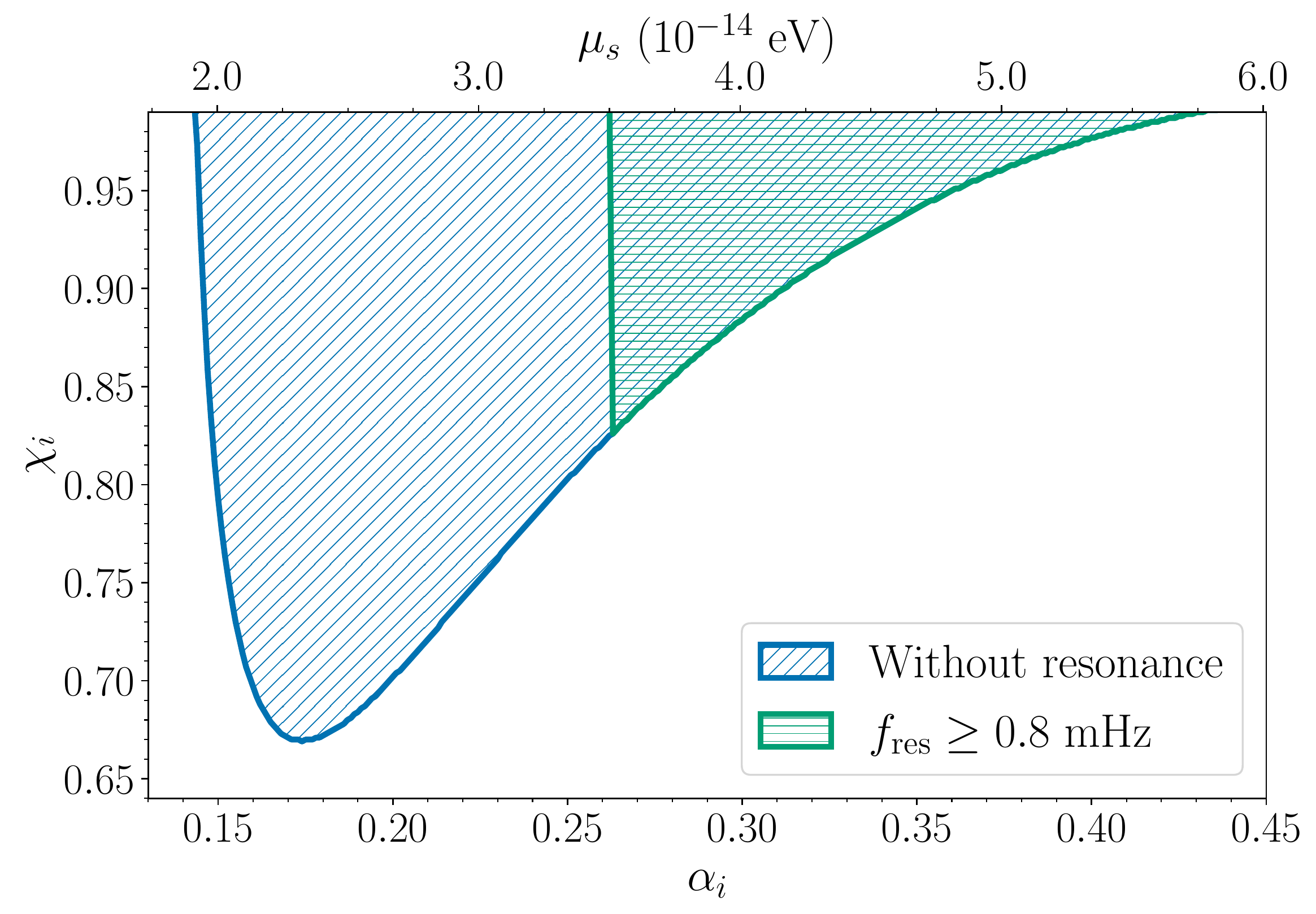}}
\caption{
Detectable parameter space $(\alpha_i,\chiIni)$ at $\DL=400$~Mpc for two example \mIni's.
The blue hatched region is the detectable space without considering orbital resonances and is identical to the blue hatched region in Fig.~\ref{fig:exc_age}.
The green hatched region corresponds to the detectable space that remains after imposing $\fres\geq\fIniLISA$, where $\fIniLISA$ is the minimum initial frequency that yields $\rhoLISA=8$:
$\fIniLISA=5.8$~mHz (0.8~mHz) for $\mIni=50\,\Msun$ ($1000\,\Msun$).
As $\fres$ increases almost monotonically with $\alpha_i$, the detectable region is truncated sharply at $\alpha_i\approx0.25$.
}
\label{fig:exc_res}
\end{figure*}

\subsection{Idealized measurement}
\label{sec:results:ideal}

Let us first assume that LISA has detected a suitable \ac{BBH} and perfectly measured the masses of its components, as well as the orbital and extrinsic parameters.
The expected GW strain produced by a boson cloud around one of the \acp{BH} will depend on the particle mass, $\mu$ (or, equivalently, $\alpha_i$), and the presuperradiance spin of the host \ac{BH}, $\chiIni$, approximately following Eq.~\eqref{eq:h0_approx}.
With knowledge of the \ac{BH} mass, we may thus chart the values of $\chiIni$ and $\alpha_i$ that would yield a detectable signal in \ac{CE}, i.e.~$h_0 \geq \hthr$.
Assuming we observe the cloud at birth, the blue region in Fig.~\ref{fig:exc_age} demonstrates this for two example \ac{BH} masses at $\DL=400$~Mpc.
For a given value of $\alpha_i$, higher $\chiIni$'s are more favorable, since those values lead to a larger initial cloud (the boson can extract more energy momentum from the \ac{BH}), and hence a stronger \ac{CW} signal.
The boundary at large $\alpha_i$'s can be understood from Eq.~\eqref{eq:final_spin}, which sets the critical spins; the boundary at small $\alpha_i$'s comes from the $\alpha_i^7$ scaling in Eq.~\eqref{eq:h0_approx}, which causes the signal amplitude to shrink rapidly as $\alpha_i$ decreases.

Unfortunately, it is safe to assume that LISA will not observe a boson cloud at birth, since simulations of BBH populations show that the typical inspiral timescale is longer than $\mathcal{O}(\rm Myr)$ before entering the LISA band~\cite{Dominik:2012kk, Mapelli:2017hqk, Mapelli:2019bnp, Neijssel:2019irh,Benacquista:2011kv, Rodriguez:2017pec, Kremer:2019iul, Antonini:2019ulv}.
Rather, the observation will occur some (long) time $\Tage$ after superradiance has taken place.
Since the GW amplitude decreases with time following Eq.~\eqref{eq:h0_evolution}, this will have a strong impact on detectability.
For $\Tage \gg \tauGW$, the observed strain amplitude at $t=\Tage$ is $h(\Tage) \approx h_0\tauGW/\Tage$ from Eq.~\eqref{eq:h0_evolution}, ignoring the frequency drift.
In the limit of $\alpha_i \ll 1$, combining Eqs.~\eqref{eq:h0_approx} and~\eqref{eq:tauGW_approx} gives a scaling relation $h(\Tage)\propto \alpha_i^{-8}$.
Hence, the strain amplitude in the large-$\alpha_i$ region is unlikely to be detected above the threshold if the \ac{BH} is too old.
In Fig.~\ref{fig:exc_age}, we use different colors to show the detectable region for different values of \Tage.
As expected, the higher $\alpha_i$ region is most affected by the age of the cloud.

Besides age, we should consider the effect of orbital resonances.
If the orbit hits a resonant frequency, cloud depletion may truncate the \ac{CW} signal during the ground-based observation, or possibly even before the binary would be detected at all.
Therefore, we must require that the initial observation frequency in LISA, $\fIniLISA$, satisfy
\beq \label{eq:fini_fres}
\fIniLISA \leq \fres\, ,
\eeq
for a possible \ac{CW} observation.
If $\fIniLISA>\fres$, the cloud vanishes due to resonant decay before the inspiral enters the LISA band and multibanding will be impossible.
The inequality Eq.~\eqref{eq:fini_fres} is not strictly necessary if there is archival \ac{CE} data before the start of LISA's observation.
However, the current timeline suggests that \ac{CE} will be online after the launch of LISA~\cite{Reitze:2019iox}.
Therefore we require Eq.~\eqref{eq:fini_fres} to be conservative.
From Eq.~\eqref{eq:fres_approx}, the resonance frequency scales as $\alpha_i^7$ for a fixed \ac{BH} mass.
This implies that resonant depletion happens at an earlier stage in the inspiral and is more likely to hinder the \ac{CE} observation in the small-$\alpha_i$ region.
The inspiral frequency at the time of the initial observation by LISA, $\fIniLISA$, is largely arbitrary, since each \ac{BBH} system is formed at a different reference time and initial separation.
Even if $\fIniLISA<\fres$, we may still miss the \ac{CW} signal: because the \ac{SNR} of a \ac{CW} signal scales as $T_{\rm obs}^{1/4}$ for a semicoherent search, the truncation may prevent \ac{CE} from accumulating enough \ac{SNR} to reach a detectable level.
(The possibility would remain, however, of observing the \ac{CW} signal in archival ground-based GW data, assuming the resonance depletion occurred within a timescale where such archival data exists.)

In Fig.~\ref{fig:exc_res}, we show the detectable region for component masses of $50\,\Msun$ and $1000\,\Msun$ at $\DL=400$~Mpc, accounting for resonances.
For systems with the parameters within the blue-hatched region but outside the green-hatched region, the boson clouds would totally deplete before the inspiral signal could be observed by LISA.
In contrast, systems with parameters enclosed by the green-hatched region do not experience resonance depletion; in this case, both the \ac{CW} and inspiral emissions are observable in \ac{CE} and LISA, respectively.
We notice that there is a sharp cutoff in the small $\alpha_i$ region when resonant depletion turns on.
This is because $\fres$ has a strong dependence of $\alpha_i^7$ as shown in Eq.~\eqref{eq:fres_approx} which suggests that $\fres$ depends on $\alpha_i$ almost exclusively.
Therefore, the criterion of $\fres\geq\fIniLISA$ prevents the observation of \ac{CW} emission from systems with small $\alpha_i$'s, irrespective of $\chiIni$.

\begin{figure*}[tb]
\centering
	\subfloat[Distance uncertainty, true $\mIni=50\Msun$.\label{fig:exc_distA}]{\includegraphics[width=0.45\textwidth]{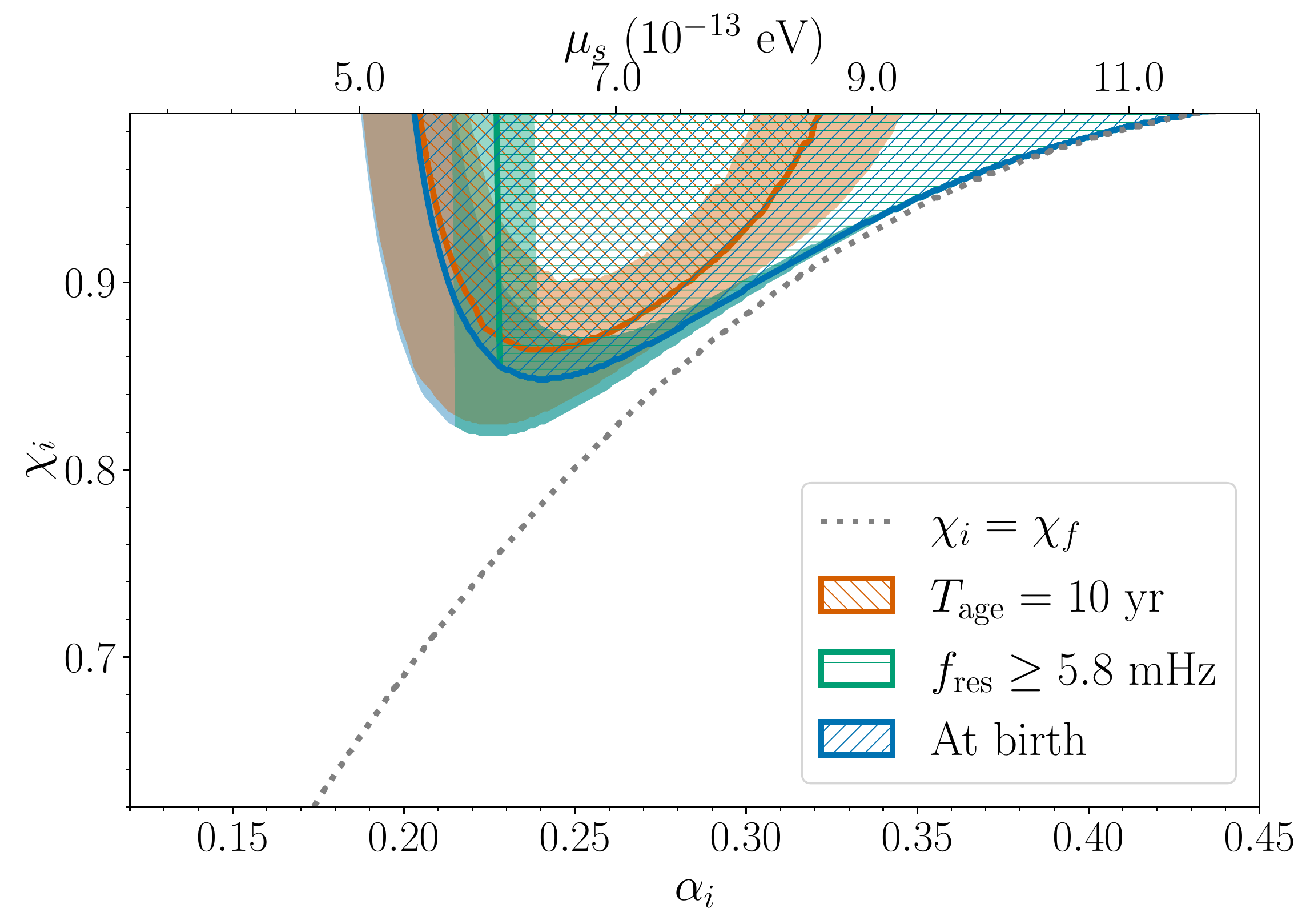}}
	\subfloat[Distance uncertainty, true $\mIni=1000\Msun$.\label{fig:exc_distB}]{\includegraphics[width=0.45\textwidth]{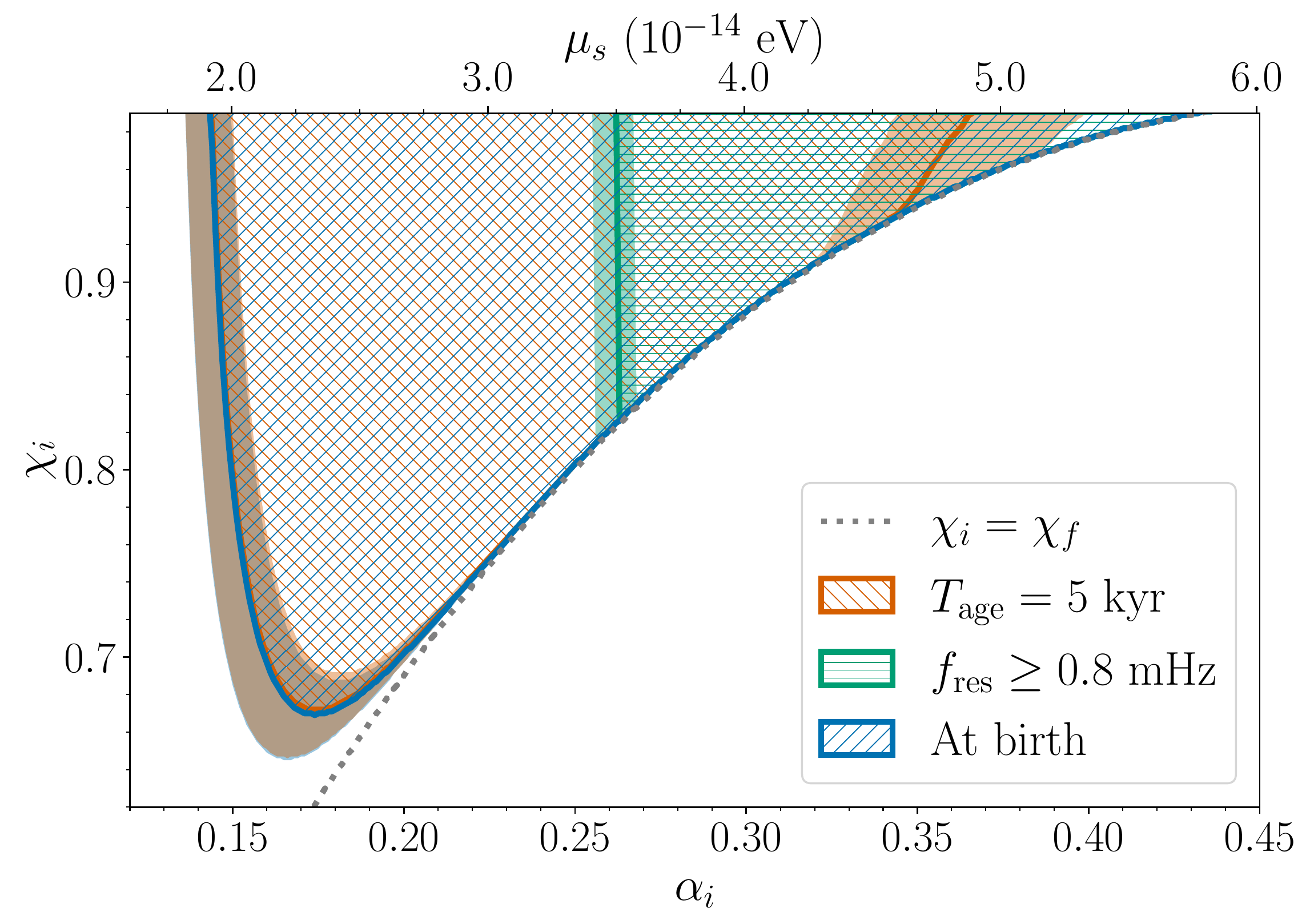}}
	\quad
	\subfloat[Mass uncertainty, true $\mIni=50\Msun$.\label{fig:exc_massA}]{\includegraphics[width=0.45\textwidth]{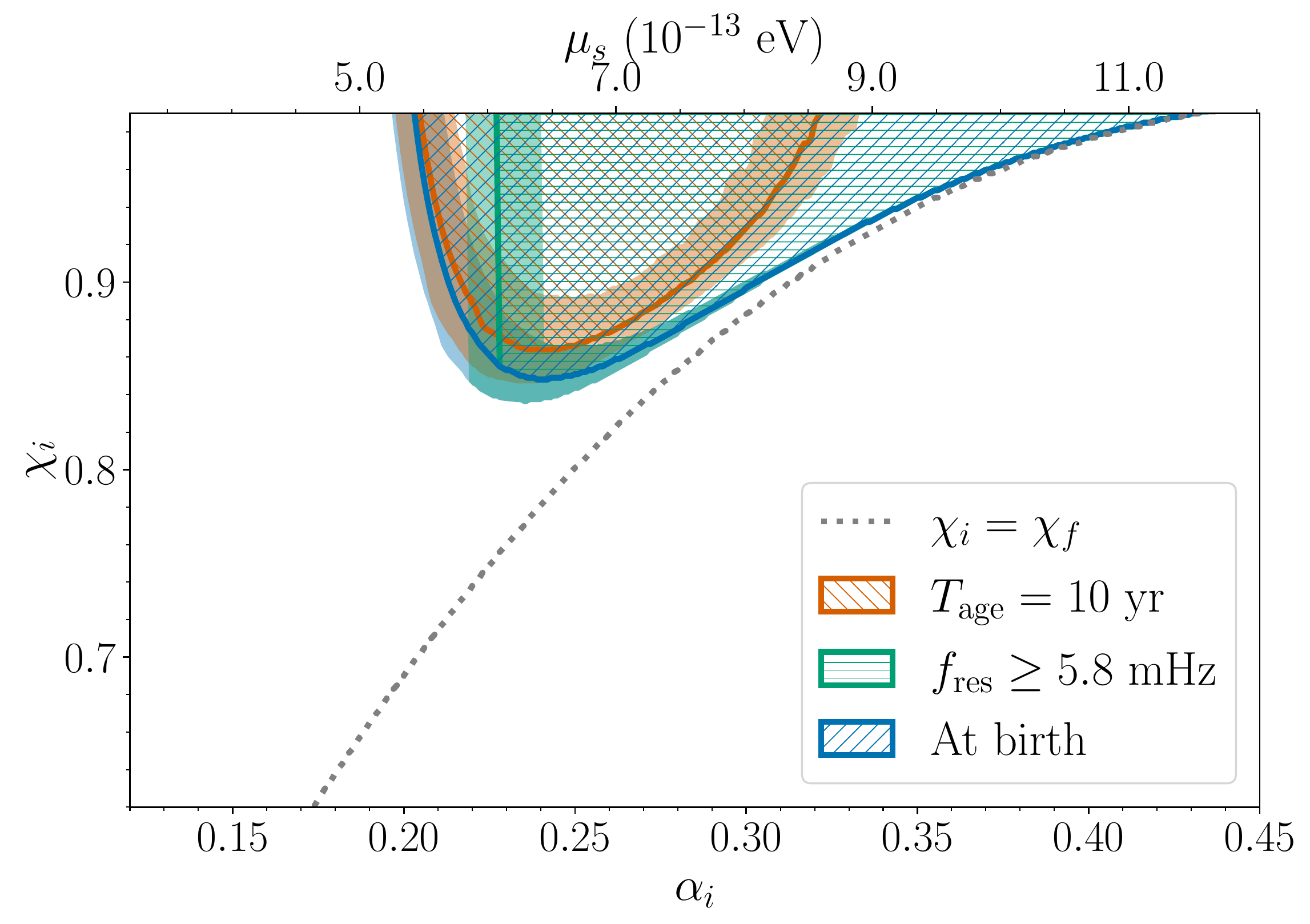}}
	\subfloat[Mass uncertainty, true $\mIni=1000\Msun$.\label{fig:exc_massB}]{\includegraphics[width=0.45\textwidth]{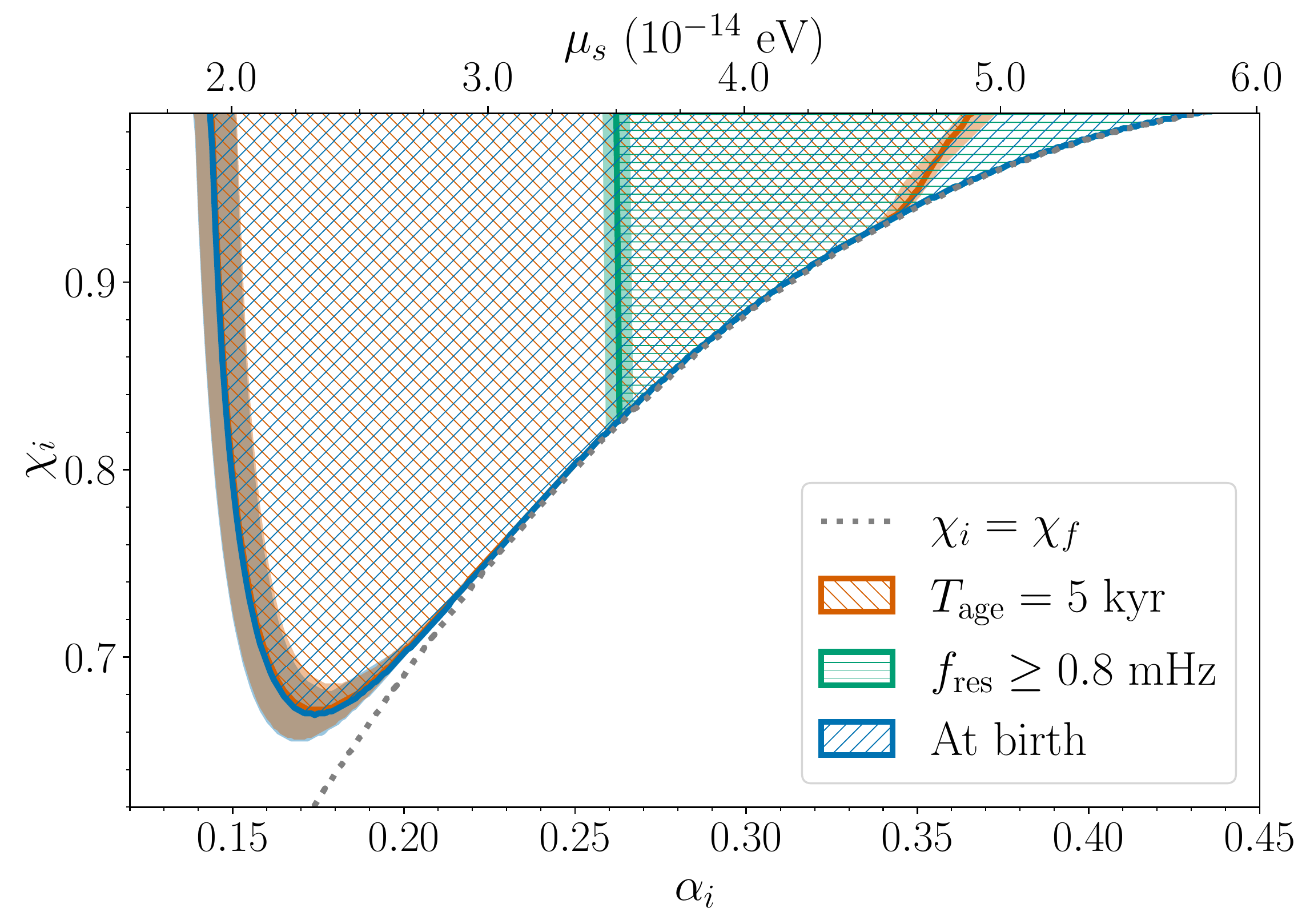}}
	\quad
\caption{
Impact of LISA measurement uncertainties on the detectable parameter space $(\alpha_i, \chi_i)$ for two example $M_i$'s, with true $D_L = 400$~Mpc and face-on inclination.
In each panel, the blue (cf.~Fig.~\ref{fig:exc_age}), orange (cf.~Fig.~\ref{fig:exc_age}), and green (cf.~Fig.~\ref{fig:exc_res}) hatched regions bounded by solid lines correspond to the detectable parameter space \textit{without resonant depletion or cloud dissipation}, \textit{with cloud dissipation only}, and \textit{with resonant depletion only}, respectively.
(The assumed $\Tage$ and $\fIniLISA$ are the same as in Figs.~\ref{fig:exc_age} and \ref{fig:exc_res}.)
The regions hatched with all colors mark parameters that remain detectable after accounting for both cloud age and orbital resonances.
The semitransparent color bands around the edge of the hatched regions encode the variation in the boundary of the inferred detectable space due to LISA measurement uncertainty in the distance (panels (a) and (b)), or mass (panels (c) and (d));
the width of the bands corresponds to the LISA 90\%-credible interval.
When the detectable regions approach the $\chi_i < \chi_f$ constraint (dotted line), the error bands shrink until disappearing; this is because $\chi_f$ is independent of $M_i$ and $D_L$, and hence uncertainty in those parameters does not affect our conclusions about detectability.
The vertical light-green band around the true $\fIniLISA$ corresponds to variations in the \emph{inferred} $\fIniLISA$ linked to the $M_i$ or $D_L$ uncertainty.
For $M_i=1000\,\Msun$, the light-blue and light-orange bands overlap almost completely at small $\alpha_i$'s, because $\tauGW \gg \Tage$ in this region.
}
\label{fig:exc_uncertainty}
\end{figure*}

\subsection{Measurement uncertainties}
\label{sec:results:uncertain}

We have, so far, assumed that LISA can provide perfect measurements of the luminosity distance $\DL$, inclination $\iota$, and \ac{BH} mass $\mObs$.
However, true LISA measurements will have uncertainties that will impact the \ac{CE} follow-up.
For example, an underestimated distance would lead to an overestimation of the \ac{CW} detectable region and, correspondingly, the range of boson masses $\mu$ that can be probed.

\begin{figure*}[tb]
\centering
	\subfloat[$\mIni=50\Msun$.]{\includegraphics[width=0.45\textwidth]{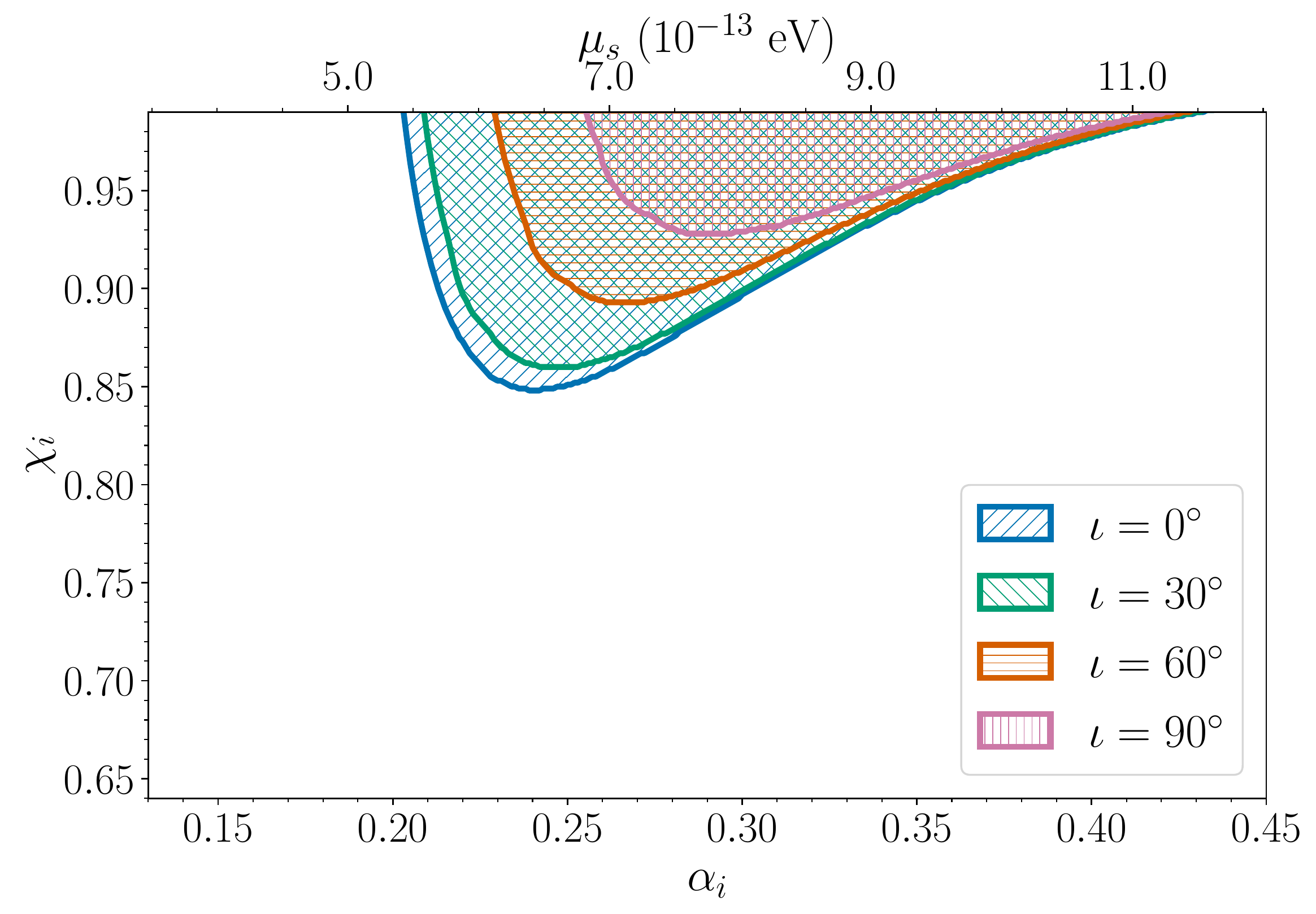}}
	\quad
	\subfloat[$\mIni=1000\Msun$.]{\includegraphics[width=0.45\textwidth]{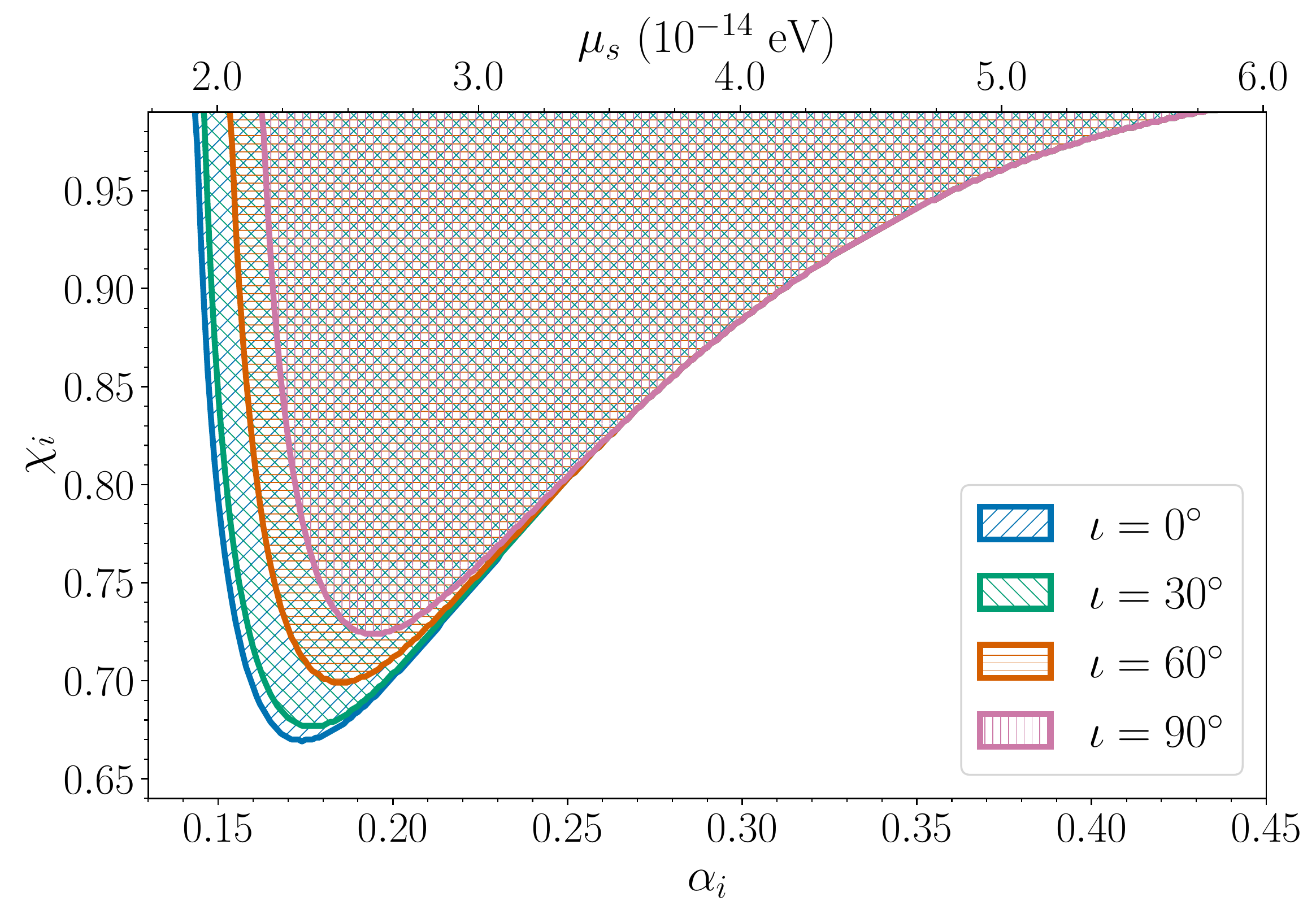}}
\caption{
Detectable parameter space  $(\alpha_i, \chi_i)$ for a \ac{BH} with different inclination angles $\iota=0^{\circ}$ (blue), $30^{\circ}$ (green), $60^{\circ}$ (orange) and $90^{\circ}$ (pink), for two examples \ac{BH} masses $M_i$, and $\DL$ again fixed at 400~Mpc.
The strain amplitude is weaker when a source is seen edge on ($\iota = 90^{\circ}$), leading the detectable parameter space to shrink as $\iota$ increases.
}
\label{fig:exc_iota}
\end{figure*}

To map how LISA uncertainties affect detectability, we calculate the detectable region corresponding to the boundaries of the projected $90\%$-credible intervals for $\mIni$ and $\DL$.
To simulate such a measurement, we assume the LISA posteriors are well represented by independent Gaussians with standard deviations given as fractions of the true values, $\sigma_{\mIni} /\mIni^{\rm true} = 0.1 $ and $\sigma_{\DL}/\DL^{\rm true} = 0.2$ for the mass and distance respectively (see Sec.~\ref{sec:lisa}).

We represent the effect of measurement uncertainty in Fig.~\ref{fig:exc_uncertainty}.
For reference, the solid blue curve encloses the parameters that we would know accessible to \ac{CE} if we had a perfect measurement of $\mIni$ and $\DL$ from LISA, assuming $\Tage=0$.
Similarly, the solid orange (green) curve encloses parameters that are accessible to \ac{CE} for a perfect LISA measurement, assuming only cloud dissipation (only resonant depletion) is present.
Since $\Tage\gg\TLISA$, we expect only the systems in the overlap region within both the resonance and age boundaries to be suitable sources.

In Figs.~\ref{fig:exc_distA} and~\ref{fig:exc_distB}, we show how such the uncertainty in the LISA distance measurement affects the expected detectable region for two example \ac{BH} masses: $\mIni=50\,\Msun$ and $1000\,\Msun$.
Each light-colored band surrounding the solid curve of the same color shows the variation of the boundary corresponding to the measurement uncertainty in $\DL$, for a fixed $\mIni$.
We compute this from the projected 90\%-credible interval as described above: the upper (lower) bound of the light-blue band corresponds to the 95\ts{th} (5\ts{th}) percentile of the $\DL$ posterior for a \ac{BH} with zero $\Tage$.
The light-orange band reflects the same projection of 90\%-credible interval of $\DL$ posterior, but $\Tage$ increases to 10 (5000) years for $\mIni=50\,\Msun$ ($1000,\Msun$).
Since the strain amplitude scales as the inverse distance, the inferred detectable region is larger for a smaller distance measurement.

On the other hand, the right (left) bound of the light-green band corresponds to the 95\ts{th} (5\ts{th}) percentile of the $\DL$ posterior.
This correspondence comes from the general correlations and interdependencies between $\DL$, $\fIniLISA$ and $\rhoLISA$.
For a constant $\rhoLISA=8$, the \textit{inferred} $\fIniLISA$ increases with larger $\DL$.
If $\fIniLISA$ is kept fixed in the analysis, $\rhoLISA$ decreases with larger $\DL$.
Finally, for sources at a constant $\DL$, $\rhoLISA$ increases with larger $\fIniLISA$ until $\fIniLISA \sim 0.1$ Hz, after which the inspiral falls out the sensitive band of LISA.
For $\fIniLISA$ greater than this bound, it is not possible to accumulate enough $\rhoLISA$ in the available observation time of the LISA mission to claim a detection.

By the same token, Figs.~\ref{fig:exc_massA} and~\ref{fig:exc_massB} shows the variation of the expected detectable regions due to the uncertainty in LISA mass measurement.
The upper (lower) bound of the light-blue and light-orange band corresponds to 5\ts{th} (95\ts{th}) percentile of the $\mIni$ posterior.
As the boson cloud will extract more energy and momentum from heavier \acp{BH}, the strain amplitude increases with the \ac{BH} mass, and the detectable region expands accordingly if we allow for higher \ac{BH} masses.
The right (left) bound of the light-green band corresponds to 5\ts{th} (95\ts{th}) percentile of the $\mIni$ posterior, since the resonance frequency is higher for smaller mass.

Generally speaking, both uncertainties on $\mIni$ and $\DL$ have a similar effect on the inferred detectable region.
The ignorance of the uncertainties on $\mIni$ and $\DL$ does not matter for the boundary of $\chi_f$ which only depends on $\chi_f$.
However, this will affect the inference of $\mu_s$ since $\mu_s \propto \alpha_i/\mIni$.
We also note that the variation does not exceed the boundary set by the final spin at the saturation of superradiance in Eq.~\eqref{eq:final_spin}, since the superradiance condition is not satisfied for $\chiIni<\chi_f$.
Due to the sharp cutoff by the resonance boundary at low $\alpha_i$, we conclude that the uncertainties on $\mIni$ and $\DL$ have a smaller impact on the detectable region compared \ac{BH} aging and resonant depletion.

Finally, we consider the effect of inclination, which we have so far assumed to be optimal.
In general, the effect of orbital precession could be included in the inspiral waveform to obtain the actual cloud's inclination $\iota$ at any given time, apart from the orbital inclination $\iota_{\rm orb}$.
Here, we consider the simpler spin-aligned case in which $\iota_{\rm orb}=\iota$ and we can directly take the LISA $\iota_{\rm orb}$ measurement as a measurement of $\iota$.
The observed polarization content varies with the inclination angle with respect to the line of sight.
For an interferometer with an interarm angle of $90^{\circ}$, such as \ac{CE}, the observed amplitude at some inclination angle $\iota$ relative to a face-on emission scales as
\begin{align}
h(\iota) \propto \sqrt{\left(1+\cos^2{\iota}\right)^2F_+^2+4\cos^2{\iota}F_{\times}^2}\, , 
\end{align}
from the quadrature addition of Eqs.~\eqref{eq:plus} and \eqref{eq:cross} weighed by the antenna pattern $F_+$ and $F_{\times}$ defined in Eq.~(57) of Ref.~\cite{Sathyaprakash:2009xs}.
$F_+$ and $F_{\times}$ depend on the zenith and azimuthal angles $(\theta, \phi)$ of the source's sky location, as well as the polarization angle $\psi$.
The inclination angle then affects the detectable region, as shown in Fig.~{\ref{fig:exc_iota}}.
To simplify the relation, we chose the source location such that $F_+=F_{\times}$.
Regardless of intrinsic parameters, the expected amplitude is the largest for face-on ($\iota=0^{\circ}$) emission, and the smallest for edge-on ($\iota=90^{\circ}$) emission.
Hence the detectable region shrinks gradually from $\iota=0^{\circ}$ to $\iota = 90^{\circ}$.

\begin{figure*}[tb]
\centering
\subfloat[Without resonance]{\includegraphics[width=0.45\textwidth]{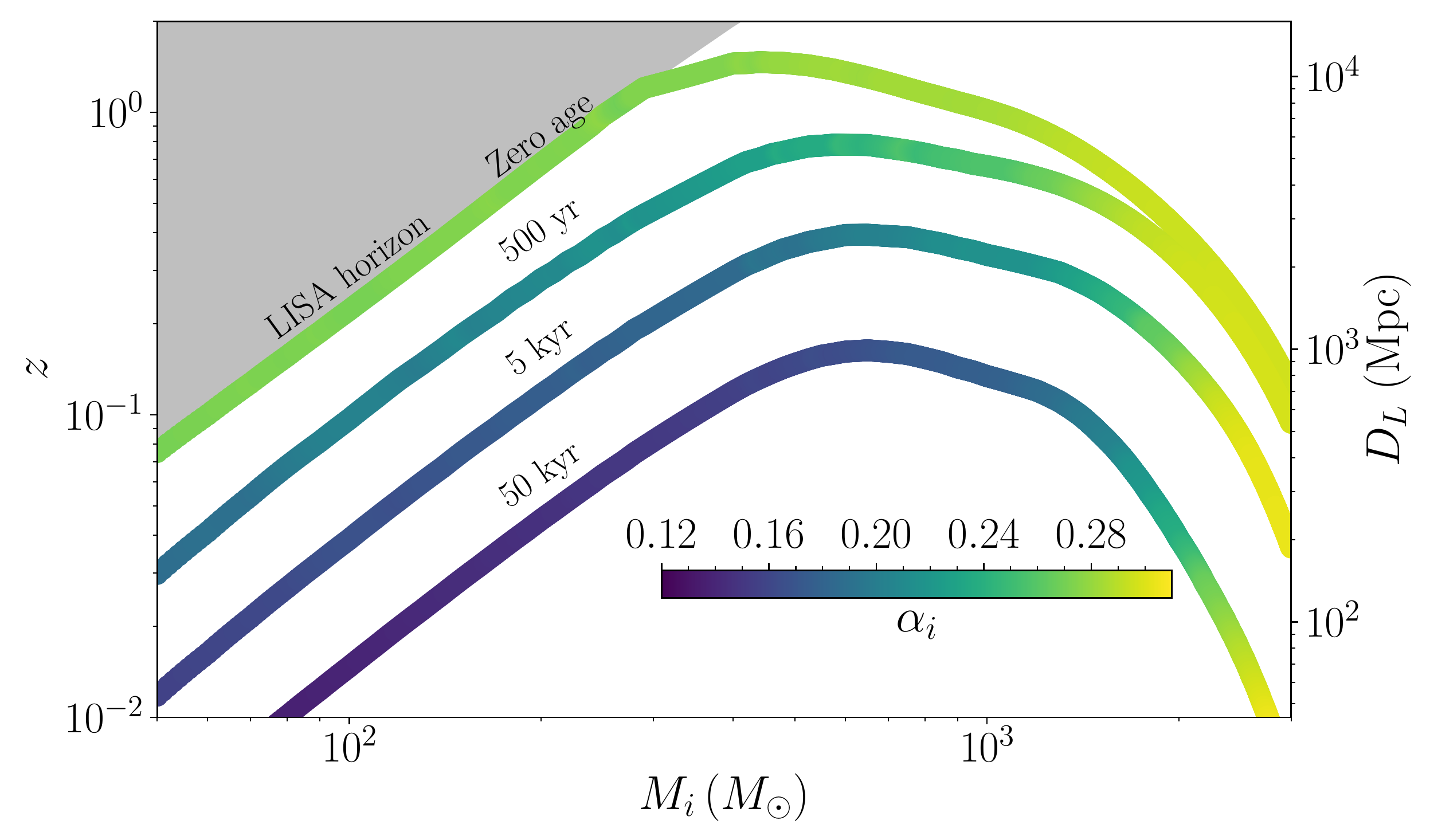}\label{fig:zHorizonNoRes}}
\subfloat[With resonance]{\includegraphics[width=0.45\textwidth]{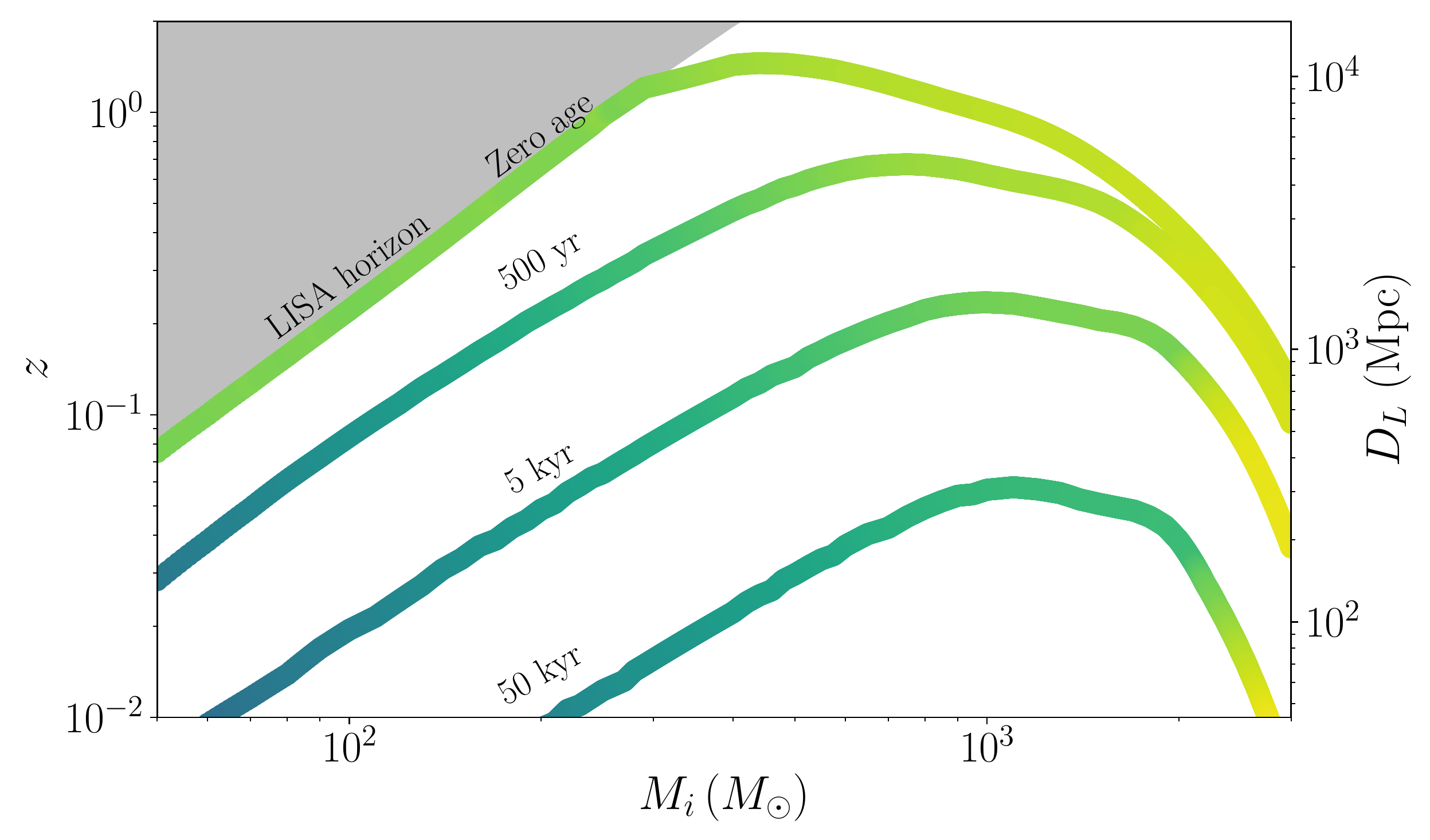}\label{fig:zHorizonRes}}
\caption{
Detection horizon as a function of $\mIni$ for $\chiIni=0.9$ and for various $\Tage$'s, as indicated next to each curve.
Color encodes the best $\alpha_i$ as described in the Sec.~\ref{sec:results:limits}.
Equal-mass \acp{BBH} inside the gray region are not detectable by LISA ($\rhoLISA<8$).
Panel (a) shows the horizons when resonant depletion is neglected, while panel (b) includes both the effects of resonant depletion and cloud dissipation.
Orbital resonances have the general effect of reducing horizons.
}
\label{fig:zHorizon}
\end{figure*}

\subsection{Detection horizon}
\label{sec:results:limits}

In the investigations above, we held the source distance fixed.
The amplitudes of GW emission from the cloud and the binary both decrease as the distance increases.
Therefore, we expect that the mostly overlapped regions in Fig.~\ref{fig:exc_uncertainty} shrink until no system can be observed.
To quantify the reach of our technique, we now investigate the detection horizon as a function of $\mIni$ and $\Tage$.
We define the detection horizon to be the largest distance such that both SNR thresholds, of \ac{CE} and LISA, are satisfied.
Thus, the detection horizon is a measure of the volume of space within the reach of our analysis technique.
For a given expected rate of \ac{BBH} mergers as a function of redshift, this quantity also informs us about the number of systems that we may expect to detect during a fixed observation period.
Unfortunately, there is large uncertainty in rates for \ac{BBH} mergers in the mass range of interest \cite{Greene:2019vlv,Salemi:2019ovz}, so we do not attempt to compute a number of expected detections.

We compute horizons assuming $\chiIni=0.9$, and assume the same definition of $\fIniLISA$ as in Sec.~\ref{sec:lisa}.
For each \ac{BH}, we identify the boson mass that generates the loudest \ac{CW}.
This results in an \textit{optimal} $\alpha_i$ that maximizes the horizon for each $\mIni$.
To include the amplitude decay due to \ac{BH} aging, we calculate the horizons for four ages: $\Tage=0$, 500 yr, 5~kyr and 50~kyr.

Figures~\ref{fig:zHorizonNoRes} and~\ref{fig:zHorizonRes} show the detection horizon as a function of $\mIni$ without and with resonant depletion, respectively.
The color scale shows the value of $\alpha_i$ that yielded maximum strain amplitude for each $(\mIni,\DL,\Tage)$.
There are some features shared by both figures.
First, since we require both detections in LISA and \ac{CE}, irrespective of resonances, systems with $\mIni \lesssim 500\Msun$ are limited by the LISA \ac{SNR} threshold if $\Tage\sim 0$, because the \ac{CE} horizon lies outside the LISA one (grey region in Fig.~\ref{fig:zHorizon}) for those systems;
otherwise, our observation is limited by the \ac{CE} \ac{SNR} threshold.
Second, all curves follow a similar shape, reflecting  the features of \ac{CE}'s \ac{PSD} (cf. Fig.~\ref{fig:cartoon}):
for $\mIni\lesssim 500\,\Msun$ ($\fCW\gtrsim20$~Hz) \ac{CE}'s sensitivity is roughly constant and, because $h_0\propto \mIni/D_L$, the detection horizon increases with $\mIni$;
then, as $\mIni$ increases further, $\fCW$ falls out of the \ac{CE} frequency band, and the detection horizon decreases.
Third, and last, $\alpha_i$ increases with $\mIni$ in both panels: this is because \ac{CE} is only sensitive to \ac{CW} frequencies ${\sim}[10,300]$~Hz, and so $\alpha_i \propto \mIni$ to stay within that range following Eq.~\eqref{eq:fgw_approx}.

There are also some features unique to both Figs.~\ref{fig:zHorizonNoRes} and~\ref{fig:zHorizonRes}, which reflect the impact of resonances on detectability.
The horizons are generally closer and the overall scale of $\alpha_i$ is higher in Fig.~\ref{fig:zHorizonRes}.
This is because systems with the true optimal $\alpha_i$ experience resonant depletion before entering the LISA band, leaving us only with suboptimal configurations.
As shown in Fig.~\ref{fig:h_alpha}, resonances prevent us from observing clouds with the overall optimal $\alpha_i$'s (crosses) for $\Tage\gtrsim1$~kyr; the best among the leftover $\alpha_i$'s (squares) lead to weaker signals.

\section{Interpretation}
\label{sec:interpretation}

The results of the \ac{CW} search must be translated into a statement about boson masses.
Doing this is straightforward if a signal is indeed found: in that case, we would be able to establish that an ultralight boson exists, and accurately infer its mass from a measurement of the \ac{CW} frequency.
Detailed tracking of the frequency and amplitude evolution would allow us to study the depletion of the cloud, infer its age, and potentially look for evidence of boson self-interactions.
The \ac{CE} measurement would also provide us with estimates of the \ac{BH} spin before and after superradiance, complementing information provided by LISA.
After establishing that at least one \ac{CW} signal is present, the information gained could be used in a targeted search to determine whether both \acp{BH} are hosting a boson cloud each.

On the other hand, if a \ac{CW} signal is not found by \ac{CE} but an inspiral signal is observed by LISA, then we would want to cast upper limits on the strain amplitude into boson mass constraints.
The translation is hindered by our lack of knowledge about the individual histories of the targeted \acp{BH}: ignorance about the cloud age and presuperradiance \ac{BH} spin preclude a direct mapping from boson mass to expected \ac{CW} strain (see Sec.~\ref{sec:results}).
Thus, unless additional information is provided by other means, we will be limited to constraints on the $\chiIni$-$\mu$-$\Tage$ space (Figs.~\ref{fig:exc_age}--\ref{fig:exc_iota}).

A significant limiting uncertainty in the interpretation of a null \ac{CW} result in CE (with the presence of an inspiral in LISA) is the difficulty to reliably quantify $\Tage$ for any individual \ac{BH}.
The predicted lifetime of a cloud is in general expected to be many orders of magnitude shorter than the time between \ac{BH} formation in a supernova and the eventual merger of a \ac{BH} binary.
For simulated populations of \acp{BBH} formed in a galactic field, typical distributions of times between \ac{BBH} formation and merger are $\mathcal{O}(\mathrm{Gyr})$ or larger, with a small-number tail reaching down to $\mathcal{O}(\mathrm{Myr})$~\cite{Dominik:2012kk, Mapelli:2017hqk, Mapelli:2019bnp, Neijssel:2019irh}.
The majority of \acp{BBH} formed through dynamical interactions in dense stellar environments would, again from simulated populations, exhibit similar timescales~\cite{Benacquista:2011kv, Rodriguez:2017pec, Kremer:2019iul, Antonini:2019ulv}, especially for binaries ejected from their formation environments (through supernova kicks).
If such a binary is not ejected however, then the additional strong dynamical encounters to which the binary would be exposed would significantly decrease the time to merger and, thus, increase the possibility that a binary with \acp{BH} harboring boson clouds survive long enough for us to observe it.
Binaries originating in dynamical environments can also be formed with significant eccentricities, or through direct captures~\cite{Breivik:2016ddj, Haster:2016ewz, Kremer:2018cir, Zevin:2018kzq,Samsing:2019lyu}, which would decrease the time to merger compared to an equivalent quasicircular orbit, and similarly increase the potential detectability of the two GW signals.
Additionally, self-interactions of the boson field could lead to a prolonged cloud lifetime \cite{Yoshino:2012kn,Yoshino:2015nsa}.

If there is no inspiral detection in LISA, then we cannot carry out a \ac{CW} follow-up in \ac{CE}, and hence can make no statements about bosons either.

Finally, the multiband technique proposed in this paper, specifically a ground-based \ac{CW} follow-up of a space-based inspiral observation, is not limited to the search of ultralight bosons.
Another possible application may be a \ac{CW} follow-up of individual neutron stars in binaries observed by LISA to look for GWs from non-axisymmetries in their moments of inertia~\cite{Pisarski:2019vxw}.

\section{Conclusion}
\label{sec:conclusion}

Future GW detectors on the ground and in space, like \ac{CE} and LISA, will be able to work together to detect ultralight bosons with masses $25 \lesssim \mu/\left(10^{-15}\, \mathrm{eV}\right)\lesssim 500$.
In detecting \ac{BBH} inspirals, LISA will provide crucial information enabling \ac{CE} to search for continuous GWs from boson clouds hosted by the component \acp{BH} (Fig.~\ref{fig:cartoon}).

In this paper, we have laid out the detection strategy (Sec.~\ref{sec:observation}), explored the relevant parameter space (Sec.~\ref{sec:results}), and discussed the interpretation of possible measurement outcomes (Sec.~\ref{sec:interpretation}).
Focusing on dominant ($\ell=m=1$) scalar clouds, we have studied limitations on potential boson constraints imposed by ignorance about the histories of the individual systems, like their age and spin evolution (Fig.~\ref{fig:exc_age}).
We have also quantified the impact of orbital resonances, which may destroy the boson cloud before its \ac{CW} signal becomes detectable by a \ac{CE} follow-up (Fig.~\ref{fig:exc_res}).
We have shown how to take all of these factors into account, together with uncertainty in the parameters measured by LISA, in order to obtain boson mass constraints from the \ac{CE} observation (Figs.~\ref{fig:exc_uncertainty}--\ref{fig:exc_iota}).
Finally, we plot detection horizons as a function of component \ac{BH} mass (Fig.~\ref{fig:zHorizon}).
Although we focused on scalars, the conclusions can easily be extended to vectors, which lead to similar phenomenology with faster timescales and greater \ac{CW} power.

\begin{acknowledgments}
The authors thank Richard Brito for useful feedback, as well as for generously providing some of his numerical results.
The authors also thank Emanuele Berti for the fruitful discussion during the APS April Meeting 2019.
The authors are also grateful to Lilli Sun for valuable discussions.
All authors are members of the LIGO Laboratory.
LIGO was constructed by the California Institute of Technology and Massachusetts Institute of Technology with funding from the National Science Foundation and operates under cooperative agreement PHY-1764464.
K.N. and S.V. acknowledge support of the National Science Foundation through the NSF Award No. PHY-1836814.
M.I.\ is supported by NASA through the NASA Hubble Fellowship Grant No.\ HST-HF2-51410.001-A awarded by the Space Telescope Science Institute, which is operated by the Association of Universities for Research in Astronomy, Inc., for NASA, under Contract NAS5-26555.
This paper carries LIGO document number \dcc{}.
\end{acknowledgments}

\bibliography{bosons}

\begin{acronym}
\acro{PSD}[PSD]{power spectral density}
\acro{ASD}[ASD]{amplitude spectral density}
\acro{SNR}[SNR]{signal-to-noise ratio}
\acro{BH}[BH]{black hole}
\acro{BBH}[BBH]{binary black hole}
\acro{BNS}[BNS]{binary neutron star}
\acro{NS}[NS]{neutron star}
\acro{BHNS}[BHNS]{black hole--neutron star binaries}
\acro{NSBH}[NSBH]{neutron star--black hole binary}
\acro{CBC}[CBC]{compact binary coalescence}
\acro{GW}[GW]{gravitational wave}
\acro{PDF}[PDF]{probability density function}
\acro{PE}[PE]{parameter estimation}
\acro{CL}[CL]{credible level}
\acro{IFO}[IFO]{interferometer}
\acro{LIGO}[LIGO]{Laser Interferometer Gravitational Wave Observatory}
\acro{LISA}[LISA]{Laser Interferometer Space Antenna}
\acro{CE}[CE]{Cosmic Explorer}
\acro{ET}[ET]{Einstein Telescope}
\acro{CW}[CW]{continuous wave}
\end{acronym}

\end{document}